\documentclass[a4paper,12pt]{article}

\usepackage{amsfonts}
\usepackage{graphicx}
\usepackage[english,activeacute]{babel}
\usepackage{amsfonts,amssymb,amsthm,amsmath}
\usepackage{graphicx,epsfig}
\usepackage{natbib}
\usepackage{color}
\usepackage[usenames,dvipsnames]{xcolor}
\usepackage{colortbl}
\usepackage{lscape}
\usepackage{multirow}
\usepackage{float}
\usepackage{siunitx}
\usepackage[a4paper]{geometry}
\usepackage{url}
\usepackage{setspace}
 \usepackage{lineno}

\newcommand{\enegrita}{\mathbf{e}}
\newcommand{\phinegrita}{\mbox{\boldmath$\phi$}}
\newcommand{\bfi}{\mbox{\boldmath$\phi$}}

\newcommand{\pinegrita}{\mbox{\boldmath$\pi$}}

\newcommand{\bmap}{\emph{BMAP}}
\newcommand{\blambda}{\mbox{\boldmath $\lambda$}}

\newcommand{\tetanegrita}{\mbox{\boldmath$\theta$}}

\newcommand{\be}{\begin{eqnarray}}
\newcommand{\ee}{\end{eqnarray}}
\newcommand{\beq}{\begin{equation}}
\newcommand{\eeq}{\end{equation}}
\newcommand{\ea}{\end{eqnarray*}}
\newcommand{\ba}{\begin{eqnarray*}}
\newcommand{\edi}{\end{displaymath}}
\newcommand{\bdi}{\begin{displaymath}}

\newcommand{\map}{\emph{MAP}}

\newtheorem{proposition}{Proposition}
\newtheorem{remark}{Remark}

\begin{document}





\title{{Analysis of an aggregate loss model in a Markov renewal regime}} 

\author{Pepa Ram\'irez-Cobo$^a$\thanks{Corresponding author: Pepa Ram\'irez-Cobo, Departamento de Estad\'istica e Investigaci\'on Operativa, Universidad de C\'adiz. Campus Universitario de la Asunci\'on. Avenida de la Universidad s/n, 11405 Jerez de la Frontera, C\'adiz (Spain), \mbox{\emph{Phone number}: +34 956 037 890}, \emph{email: pepa.ramirez@uca.es}}, Emilio Carrizosa$^b$ and Rosa E. Lillo$^{c,d}$\\ 
{\small $^a$\emph{Departamento de Estad\'istica e Investigaci\'on Operativa, Universidad de C\'adiz (Spain)}}\\ {\small $^b$\emph{IMUS, Instituto de Matem\'aticas de la Universidad de Sevilla (Spain),}}\\ {\small $^c$\emph{Departamento de Estad\'istica, Universidad  Carlos III de Madrid (Spain)} }
\\ {\small $^d$\emph{uc3m-Santander Big Data Institute, Universidad  Carlos III de Madrid (Spain)}}
}
\maketitle

\begin{abstract}
In this article we consider an aggregate loss model with dependent losses. The loss occurrence process is governed by a two-state Markovian arrival process ($\map_2$), a {Markov renewal process that allows for (1) correlated inter-loss times,  (2) non-exponentially  distributed inter-loss times and, (3) overdisperse loss counts. Some quantities of interest to measure persistence in the loss occurrence process are obtained. Given a real OpRisk database, the aggregate loss model is estimated by fitting separately the inter-loss times and severities. The $\map_2$ is estimated via direct maximization of the likelihood function, and severities are modeled by the heavy-tailed, double-Pareto Lognormal distribution. In comparison with the fit provided by the Poisson process, the results point out that taking into account the dependence and overdispersion in the inter-loss times distribution leads to higher capital charges}.
\end{abstract}
%

\section{Introduction}\label{sec: intro}
The modeling of losses in actuarial science has occupied a prominent role in the last decades. In particular, the estimation of aggregate loss models constitutes a challenge for the insurance industry and financial entities due to the need for a correct assessment of the premium rates and safety margins. The standard definition of an aggregate loss models is as follows. Let $X_1,X_2,\ldots$ be independent and identically distributed random variables representing the sizes of individual claims (or severities), and let $N$ count for the total number of losses in a fixed period (frequency). We shall assume that the severities are independent of the occurrence process. Then, the total loss for the time period can be written as the compound model
\begin{equation}\label{risk_model}
Z = \sum_{i=1}^{N}X_i.
\end{equation}
The usual approach for defining the aggregate loss, which shall be considered in this paper, is to separately model the frequency and the severity of the losses. Then, the distribution of $Z$ is obtained from that of $N$ and the sequence $X_1,X_2,\ldots$. For a thorough description of general aggregate models, the reader is referred to \cite{PanjerChapter} or \cite{klugman2012loss}. 

One of the most significant applications of loss aggregate distributions is the modeling of the operational risk (OpRisk). The OpRisk constitutes, together with market and credit, the financial risk and results in different types of losses as those generated by natural disasters, frauds, terrorist attacks or security and legal risks. Typically, in the OpRisk modeling, the total loss is defined as in (\ref{risk_model}), an expression which can be generalized to consider different risk cells or business lines. 


{Dependence has been a key issue in the recent risk research. The classical risk model has been extended in a number of ways to obtain more general models assuming dependence between claim sizes or inter-claim times, see for example \cite{AMC6}.  Dependence has been also considered in multiple ways for OpRisk aggregate loss models.} For example, dependencies among frequencies and/or severities can exist within or between cells \citep{cope2008observed,mittnik2009estimating,double}. Also, the type of loss may be a  factor to generate dependence patterns, as for instance, in \cite{ren2012multivariate}. Other authors consider models where the correlation is introduced at the aggregate loss level, as in \cite{nystrom2002quantitative,Chapelle}. Copulas constitute a widely used approach for modeling dependence between risk cells, see \cite{Embrechts2,Brechmann,Mcneil}, to name a few among multiple works. Mixture models have also been considered in the OpRisk literature to introduce dependence among frequencies and severities, as in \cite{Reshetar08}. Finally, point processes are also one of the tools employed for constructing dependent occurrence processes, as shown in 
\cite{pfeifer2004modeling,chavez2006quantitative,fung2019multivariate}. 

{As pointed out in \cite{Chernobai1}, an alternative approach for studying the frequency of losses in a OpRisk context is to look into the statistical properties of the inter-loss times. As previously commented, the Poisson process has been widely used for modeling the occurrence of losses. However, it might be the case that the exponential distribution does not properly fit real datasets, and in such cases a more general continuous distribution supported on the positive real line is needed \citep{Chernobai1}. In addition, although few studies have been devoted to examine the dependence on time of the losses, others point towards the need of considering serial correlations \citep{cope2008observed, chavez2015extreme}. Finally, the equidispersion property of the Poisson process may be inadequate in practice: for example, in the Cruz study fraud loss data \citep{Cruz} the empirical variance doubles the mean. This overdispersion characteristic of the OpRisk has been also documented by \cite{overdispersion,overdispersion2}.}

{The contribution of this paper is three-fold. First, we consider an aggregate loss model where the losses occurrences are modeled by a Markov renewal process that allows for non-exponential and correlated inter-loss times as well as for overdisperse loss counts. Second, some measures of persistence related to the loss occurrence process which may be of interest for financial decision makers are derived. In particular, closed-form expressions for the transition probabilities between short and long inter-loss periods as for the mass probability function of  long/short inter-loss times spells are calculated. Third, the aggregate model and quantities of interest as persistence coefficients and capital charges are estimated on a base of a real OpRisk dataset.}

{Markov renewal process and renewal processes have been a traditional tool for insurance risk modeling, see for example \cite{maegebier2013valuation,fodra2015high,AMC5,AMC3}. In this work we consider the Markovian arrival process ($\map$), a type of Markov renewal process which is in turn a subtype of  Batch Markovian arrival process ($\bmap$).
The general $\bmap$ \citep{Neuts79} defines a versatile class of point process containing both renewal and non-renewal processes, as the Phase-type renewal processes and Markov modulated Poisson processes, respectively.} In a $\bmap$, the times between events (losses, arrivals...) are dependent and phase-type (PH) distributed, and as will be described later, the mean and variance of the counting process do not coincide. All these properties have made the $\bmap$ suitable for modeling a variety of real life contexts as queuing theory, reliability, teletraffic, and climatology, see  \cite{Lucantoni.new.results,AMC1,hydro,joanna2,AMC4,yoelEjor}, just to cite a few. In particular, the $\map$ has been also widely considered in the insurance literature, see \cite{ng2006joint,badescu2007analysis,ahn2007analysis,frostig2008ruin, cheung2009perturbed,Cheung,zhang2011absolute}. In \cite{ren2012multivariate}, a variant of the $\map$, the Marked Markovian arrival process, which differentiates between multiple categories of losses, is considered.


{A number of papers have considered statistical inference for different classes of $\bmap$s. For example, \cite{Telek}, \cite{Eum}, \cite{Bodrog}, \cite{Casale}, \cite{joanna2} and \cite{yoelEjor} investigate moments matching
approaches. Other authors adopt a Bayesian viewpoint, as \cite{fearnhead2006exact} or \cite{pepaBayes}, where exact Gibbs samplers are proposed to obtain the posterior distribution of the model parameters. Finally, the EM algorithm is the method suggested for instance in \cite{Breuer}, \cite{Klemm} and more recently, in \cite{okamura2016fitting}. In this paper, given a real OpRisk dataset, we carry out estimation of the aggregate loss model by fitting separately the frequency and severities. In regards the frequency, we apply direct maximization of the likelihood for a sequence of consecutive inter-loss times in a $\map_2$, an approach that follows \cite{carrizosa2014maximum}.}

As previously commented, heavy tailed distributions constitute the common tool for modeling severities. In our example, the double-Pareto Lognormal distribution \citep{Reed} provides reasonable estimates for both the body and tail of the empirical distribution. Once the occurrence process and the severities are fitted, we are able to estimate the loss aggregate distribution and  both persistence and risk related measures as the distribution of short (long) inter-loss times spells, or the Value-at-Risk and Expected Shortfall.  It should be noted that in spite of the large amount of aggregate loss models defined in the literature, papers dealing with their statistical estimation are very scarce. As examples, see \cite{dutta2006tale}, \cite{valle} and \cite{ausin2011bayesian}. In the first, a simple distribution (Poisson or Negative Binomial) is fitted to the frequency data; \cite{valle} consider Bayesian estimation for a loss model under different combinations of distributions for the pair frequency/severity. Finally, in \cite{ausin2011bayesian}, a different model (the Coxian distribution) is estimated to inter-loss data. However, in all the previous works independent losses were assumed.

{
This paper is organized as follows. Sections 2.1 and 2.2 review the definition and properties of the $\map_2$, in particular those related to the inter-loss time distribution and the loss counting process. Section 2.3 describes the estimation algorithm for fitting the $\map_2$ to a sequence of inter-loss times. In Section 3 we analyze in detail several aspects of the $\map_2$ that remained unexplored, up to our knowledge. In particular, Section 3.1 considers the issue of the overdispersion, while Section 3.2 is devoted to present novel measures of persistence in the loss occurrence process. While Sections 2 and 3 are focused on the loss occurrence process, Sections \ref{review_loss} and \ref{numerical} deal with the aggregate loss model. Section \ref{review_loss} reviews basic theoretical aspects of loss aggregate models in the context of OpRisk and Section \ref{numerical} represents the applied contribution of the paper. The inference of the $\map_2$ allows for estimating the considered aggregate loss distribution, given a real OpRisk data set. Setting a specific type of heavy-tailed distribution for the severities (double Pareto Lognormal distribution, see \cite{Reed}), then a Monte Carlo algorithm, for which the convergence is assessed, is used to simulate samples from the aggregate loss model. The differences obtained under a $\map_2$ and, the commonly assumed Poisson process, in terms of the loss distribution and risk measures are object of discussion. Finally, Section \ref{sec: discussion} considers conclusions and future work.}

{
\section{A Markov renewal loss occurrence process}
The Markovian arrival process ($\map$) constitutes a class of Markov renewal process that generalize in matrix form the Poisson process by allowing for dependent and non-exponentially distributed inter-loss times. In this paper we will concentrate on the stationary two-state $\map$, noted $\map_{2}$, for two reasons. First, it has proven to be a versatile process in a number of different contexts \citep{Heffes.packetized,scott1999bayesian,hydro,pepaBayes}. Second, unlike higher order $\map$s, it can be represented by a canonical, unique parametrization in terms of a small number of parameters, which are convenient properties from a statistical inference viewpoint. This section summarizes the main properties of the $\map_{2}$, with special emphasis on the inter-loss time distribution and the counting process.}

\subsection{Definition of the $\map_2$}
The $\map_2$ is a doubly stochastic process defined by a Markov process with state space $\mathcal{S}=\{1,2\}$ and a counting process that changes according to the transitions of the Markov process.  For a detailed description of the $\map_2$, see
\cite{Neuts79,Lucantoni90,Lucantoni,Ramirez}.

In the $\map_2$, the initial state $i_0\in \mathcal{S}$ is sampled from an initial probability vector
$\tetanegrita=(\theta,1-\theta)$. At the end of a sojourn time which follows an exponential distribution with rate $\lambda_i$, two different type of transitions may occur. First,
with probability $0\leq p_{ij1}\leq 1$ a unique loss occurs and
the $\map_2$  goes to a state $j\in \mathcal{S}$.
On the other hand, with probability $0\leq p_{ij0}\leq 1$, no loss occurs
and the $\map_{2}$ goes to a state $j$ which is necessarily different from state $i$.  

A stationary $\map_2$ can be represented by  $\{\blambda, P_{0}, P_{1}\},$  where $\blambda = (\lambda_{1},\lambda_{2})$, and $P_{0}$ and $P_{1}$ are $2\times 2$ probability matrices with elements corresponding to transition probabilities $p_{ij0}$ ($i\neq j$) and $p_{ij1}$, respectively. Instead of transition probability matrices, any $\map_2$ can also be characterized in terms of the rate matrices $\{D_{0},D_{1}\}$, 
\begin{equation}\label{known map}
D_{0}=\begin{pmatrix}
  -\lambda_1 &\ \lambda_1p_{120}  \\
  \lambda_2 p_{210} &\ -\lambda_2 \\
\end{pmatrix},\quad
D_{1}=\begin{pmatrix}
  \lambda_1p_{111} &\ \lambda_1(1-p_{120}-p_{111})  \\
  \lambda_2p_{211} &\ \lambda_2(1-p_{210}-p_{211}) \\
\end{pmatrix}.
\end{equation}
The matrix $D_{0}$ is assumed to be stable, and as a consequence, it is nonsingular and the sojourn times are finite with probability 1. The definition of $D_{0}$ and $D_{1}$ implies that $D = D_0+D_1$ is the infinitesimal generator of the underlying Markov process, with stationary probability vector $\pinegrita=(\pi,1-\pi)$, computed as $\pinegrita D = {\bf 0}$. When $D_0=-\lambda$ and $D_1 = \lambda$ the $\map_2$ is reduced to the Poisson process with rate $\lambda$.

The $\map_2$ is a Markov renewal process. Indeed, if
$Y_{n}$ denotes the state of the  $\map_2$ at the time of the $n$th
loss, and let $T_{n}$ denote the time elapsing such loss and the previous one, then $\{Y_{n-1},T_{n}\}_{n=1}^\infty$ is a Markov renewal
process. {In this case, the associated $2\times 2$ semi-Markov kernel, whose $(i,j)$-th element is defined as 
\begin{equation*}
Q(i,j,t) = P\left(Y_{n+1}=j,T_{n+1}\leq t\mid Y_n=i\right),
\end{equation*}
 is given by
 \begin{equation}\label{semiMk}
Q(t)=\left(I-e^{D_0t}\right)\left(I-P_{0}\right)^{-1}P_{1},
\end{equation}
see \cite{Chakravarthy}.}

From Markov renewal theory, $\{Y_{n}\}_{n=1}^\infty$ is a Markov
chain whose transition matrix $P^\star$ is given by
\begin{equation}\label{Pstar}
P^\star =\left(I-P_{0}\right)^{-1}P_{1},
\end{equation}
with stationary distribution $\phinegrita$ as
\begin{equation}\label{phi}
\phinegrita = (\pinegrita D_1 \enegrita)^{-1}\pinegrita D_1,
\end{equation}
see \cite{Ramirez}. It is a straightforward computation to prove that (\ref{Pstar}) can be also rewritten as $P^\star =\left(-D_0\right)^{-1}D_{1}$.

The expression (\ref{known map}) for the $\map_2$ in terms of $6$ parameters is known to be overparameterized, \cite{Ramirez}.
However, \cite{Bodrog} provide a unique, canonical representation for the $\map_{2}$ in terms of just four parameters.
Such canonical representation is the one we are using in this paper. Let $-1\leq \gamma<1$ be one of the two eigenvalues of the transition matrix $P^\star$
(since $P^\star$ is stochastic, then necessarily the other eigenvalue is equal to 1), \cite{Bodrog}. If $\gamma>0$, then the canonical form of the $\map_2$ is given by
\begin{equation}\label{can1}
D_{0}=\left(\begin{array}{cc} x & y\\
0 & u\end{array}\right), \qquad D_{1}=\left(\begin{array}{cc} -x-y & 0  \\
v & -u-v \end{array}\right).
\end{equation}
On the other hand, for those $\map_2$s such that $\gamma\leq 0$, then their canonical form is
\begin{equation}\label{can2}
D_{0}=\left(\begin{array}{cc} x &  y\\
0 & u\end{array}\right), \qquad D_{1}=\left(\begin{array}{cc} 0 & -x-y  \\
-u-v & v \end{array}\right).
\end{equation}
{In the previous representations (\ref{can1}) and (\ref{can2}), $x$ and $u$ are less than $0$ while the rest of elements are larger or equal to $0$.}


\subsection{The inter-loss times distribution and the loss counting process}
Special attention deserves the analysis of the $T$, the random variable representing the time
between two sequential losses of a $\map_2$ (or the inter-loss time distribution) in its stationary version. It is known that $T$ is phase-type (PH) distributed with representation $\{\phinegrita, D_{0}\}$ \citep{Chakravarthy}, and therefore the cumulative distribution function is 
\begin{equation}
F_{T}(t) = 1-\phinegrita e^{D_{0}t}\enegrita,\quad \text{ for }t\geq 0. \label{cdf_ph}
\end{equation}
{It should be pointed out here that the phase-type distribution generalizes both the exponential distribution as well as the Coxian distribution (used by \cite{ausin2011bayesian} in the context of aggregate loss distributions)}.

The moments of $T$ can be computed as
\begin{equation}\label{moments}
m_n=E(T^n)=n!\phinegrita \left(-D_0\right)^{-n} \enegrita,
\end{equation}
where $\phinegrita=(\phi, 1-\phi)$ is the probability distribution satisfying $\phinegrita P^\star = \phinegrita,$
and $\enegrita$ is a vector with elements equal to $1$. In particular, the loss rate is computed as
\begin{equation}\label{rate}
\lambda^{\star} = 1/m_{1} = \pinegrita D_{1}\enegrita.
\end{equation}
If the focus is not on the marginal properties of inter-loss times but on their joint structure then, the likelihood function for a $\map_2$ trace of inter-loss times is given by
\begin{equation}\label{lik}
f(t_1,t_2,\ldots,t_n|D_0,D_1) =\bfi e^{D_{0}t_{1}}D_{1} e^{D_{0}t_{2}}D_{1}\ldots e^{D_{0}t_{n}}D_{1}  \enegrita,
\end{equation}
see for example \cite{Breuer,Ryden96b}.

As commented in Section 1, the $\map_2$ allows for dependent inter-loss times. In particular, the consecutive times in the stationary version of the $\map_2$ are correlated according to the correlation coefficient given by
\begin{equation}\label{acf_map2}
\rho=\gamma\ \frac{\displaystyle\frac{m_2}{2}-m_1^2}{m_2-m_1^2},
\end{equation}
where, as commented in the previous section, $\gamma$ denotes the eigenvalue of the transition matrix $P^\star$ that satisfies  $-1\leq \gamma<1$.

Next we shall briefly describe the counting process. Let $N_t(\tau)$ represent the number of losses in the interval $[t, t+\tau]$. Then, the number of losses observed in the interval $(0,\tau]$ will be given by $N_0(\tau)$. For $n\in \mathbb{N}$ and $\tau\geq 0$, let $P(n,\tau)$ denote the $2\times 2$ matrix whose $(i,j)$-th element is 
\begin{equation*}\label{Pij}
P_{ij}(n,\tau)=P\left(N_0(\tau)=n,\ J(\tau)=j \mid N_0(0)=0,\ J(0)=i\right),
\end{equation*}
for $1\leq i,j \leq 2$. From the previous definition it is clear that
\begin{equation}\label{PN}
P\left(N_0(\tau)=n\mid N_0(0)=0\right)=\pinegrita P(n,\tau) \enegrita.
\end{equation}

The values of the matrices $P(n,\tau)$ cannot be computed in closed-form. Their numerical computation is based on the \emph{uniformization method} addressed in \cite{NeutsLi} and summarized for the case of the $\map_2$ in \cite{joanna2}.

%

\subsection{Statistical estimation of the $\map_2$}\label{inference2}
A number of papers has considered estimation for different classes of $\bmap$s, the method of moments and the EM algorithm being the most studied approaches, see \cite{Bodrog,Casale,joanna2,yoelEjor,Breuer,okamura2016fitting}. Given a real trace of inter-loss times from the OpRisk context, we explore in this paper the estimation of the $\map_2$ via a direct maximization of the likelihood function in the same spirit as in \cite{carrizosa2014maximum}. In particular, given a sequence of inter-loss times ${\bf t}=(t_1,t_2,\ldots,t_n)$, we formulate the ML estimation problem from the likelihood expression (\ref{lik}) as   

\begin{equation*}(P1)\label{problema}
\left\{
\begin{array}{lll}
\max & \bfi e^{D_{0}t_{1}}D_{1} e^{D_{0}t_{2}}D_{1}\ldots e^{D_{0}t_{n}}D_{1}\enegrita   \\
\\
\mbox{s.t.} &  D_{0}=\left(\begin{array}{cc} x & y\\
0 & u\end{array}\right), \\ \\
 & D_{1}=\left(\begin{array}{cc} -x-y & 0  \\
v & -u-v \end{array}\right),\\ \\
 &x,\ u,\ x+y,\ u+v\leq 0,\\
 &y,v \geq 0,\\
 & \phinegrita (-D_0)^{-1}D_1 = \phinegrita.
\end{array} \right.
\end{equation*}
Note that $(P1)$ is stated in terms of the first canonical form of the $\map_2$, see (\ref{can1}). Regarding the second canonical form, we formulate $(P2)$ as $(P1)$, substituting the expressions for $D_0$ and $D_1$ by (\ref{can2}). In practice, given ${\bf t}=(t_1,t_2,\ldots,t_n)$, both problems $(P1)$ and $(P2)$ are solved and then the solution that maximizes the likelihood is kept.

{Some computational aspects are detailed next.} In order to solve $(P1)$ (or $(P2)$), we used the MATLAB${}^\copyright$ routine {\tt fmincon} (with default options), where the starting solution was selected as follows. It is known that the $\map_2$ is completely characterized by the first three inter-loss times moments $m_1$, $m_2$, $m_3$ as in (\ref{moments}), and by the correlation coefficient $\rho$ as in (\ref{acf_map2}), see \cite{Bodrog}. Then, a moments matching problem can be defined as
\begin{equation*}(P0)
\left\{
\begin{array}{lll}
\min & \displaystyle \delta(x,y,u,v) \\
\mbox{s.t.} & y,v\geq 0,\\
 & x,u\leq 0, \\
 &x+y\leq 0,\\
 &u+v\leq 0,
\end{array}
\right.
\end{equation*}
where $\delta(x,y,u,v)$ is the objective function that  measures the distance between the theoretical and empirical moments ($\bar{m}_1$, $\bar{m}_2$, $\bar{m}_3$ and $\bar{\rho}$):
\begin{eqnarray*}\label{objective}
\!\!\!\!\!\!\delta(x,y,u,v) &=& \left\{\rho(x,y,u,v)-\bar{\rho} \right\}^2 +\\ \nonumber
&+&
\left(\frac{m_1(x,y,u,v)-\bar{m}_1}{\bar{m}_1}\right)^2   + \left(\frac{m_2(x,y,u,v)-\bar{m}_2}{\bar{m}_2}\right)^2 +\left(\frac{m_3(x,y,u,v)-\bar{m}_3}{\bar{m}_3}\right)^2.
\end{eqnarray*}
{To solve the multimodal Problem (P0), the
 routine {\tt fmincon} was also used. A  multistart was then executed with $100$ randomly
chosen starting points and found to yield satisfactory
results. 
In practice, it turned out that the solution to the moments matching method is a good starting point for the ML problem. Indeed, other initial values could have been chosen, for example random starting $\map_2$s. In order to analyze the effect of the different starting points on the performance of the direct maximization likelihood approach, a 
total of one thousand random $\map_2$s were estimated as the solution of problem (P1), where the starting values were (1) randomly generated versus (2) the moments matching estimates. It was found that in the $61.53\%$ of times, the objective
function (the likelihood function) was larger using a moments method estimate. Finally, we discuss about the computational times characterizing the inference approach. Solving problem (P0) is fast in practice since it does not depend on samples sizes (the input is not the sequence of inter-loss times but the four empirical moments $\bar{m}_1$, $\bar{m}_2$, $\bar{m}_3$ and $\bar{\rho}$). For a prototype code written in MATLAB${}^\copyright$ the running time was equal to $6.3$ seconds, when run on Intel Core i5 at 3.2 GHz and 16 GB of DDR3 RAM. Solving (P1) is slower than solving (P0), since it does depend on samples sizes. Using the solution of (P0) as the starting solution and a sample of size equal to $n=100$, the running time was $18.2$ seconds. If instead, a sample of $n=500$ inter-loss times is considered, the running time increased to $30.7$ seconds. The undertaken approach is in any case faster than other approaches. If the Bayesian method in \cite{pepaBayes} is implemented, then for a sample of $500$ inter-loss times the running time is about $19$ seconds for a total of $1000$ iterations. Taking into account that convergence is achieved after $50000$ iterations, the Bayesian method turns out significantly slower. Recently, \cite{yoelEjor} compared the EM algorithm with a moments matching method in the context of Batch Markovian arrival processes, resulting the last one considerably faster.}

The estimation of the $\map_2$ is not the main goal of this paper, and therefore we refer the reader to \cite{carrizosa2014maximum}, where the technical details regarding the previous approach as well as several numerical illustrations are discussed more in depth.



%
%

\section{Overdispersion and persistence}

\subsection{Overdispersion under the $\map_2$}
As commented in Section 1, there are works pointing towards the overdispersion of the loss counts in the OpRisk context, which implies that the mean of the cumulated losses is considerably smaller than its variance, see \cite{Cruz,overdispersion,overdispersion2}. Unlike the Poisson process, the $\map_2$ is able to model such overdispersion phenomenon, as we illustrate next.

Throughout this work, $N(\tau)$ shall denote the stationary counting process in an interval of length $\tau$, that is $N(\tau)=\lim_{t\rightarrow \infty}N_t(\tau)$ where $N_t(\tau)$ counts for the number of losses in the interval $[t, t+\tau]$. Then, the mean number of losses is defined in terms of the loss rate (\ref{rate}) as
\begin{equation}\label{expected_losses}
E\left[N(\tau)\right]=\lambda^{\star}\tau,
\end{equation}
and the variance of that count is given by
\begin{equation}\label{variance_count}
V\left[ N(\tau)\right] = (1+2\lambda^{\star})E\left[N(\tau)\right]-2\pinegrita D_1\left(\enegrita\pinegrita +D  \right)^{-1}D_1\enegrita \tau-2\pinegrita D_1\left(I-e^{D\tau}  \right)\left(\enegrita \pinegrita+D\right)^{-2}D_1 \enegrita,
\end{equation}
see \cite{narayana} or \cite{Eum}.

\begin{figure}[h]
\begin{center}\begin{tabular}{c}
\includegraphics[height=4in]{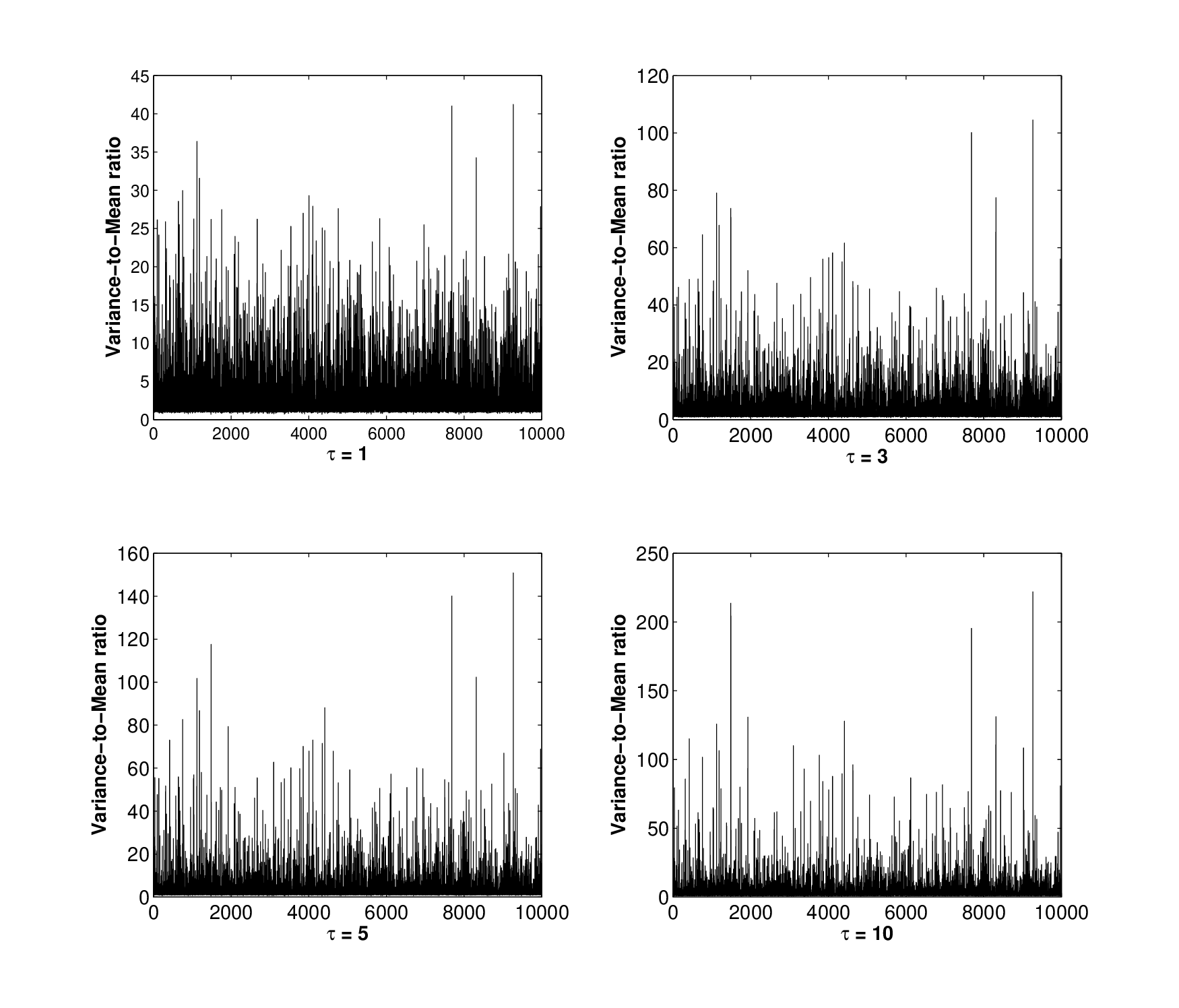}
\end{tabular}
\end{center}
\caption{\label{Fig:VtM}Variance-to-Mean ratios for a total of $10000$ simulated $\map_2$s for different values of $\tau$.}
\end{figure}

\begin{figure}[h]
\begin{center}\begin{tabular}{c}
\includegraphics[height=2.8in]{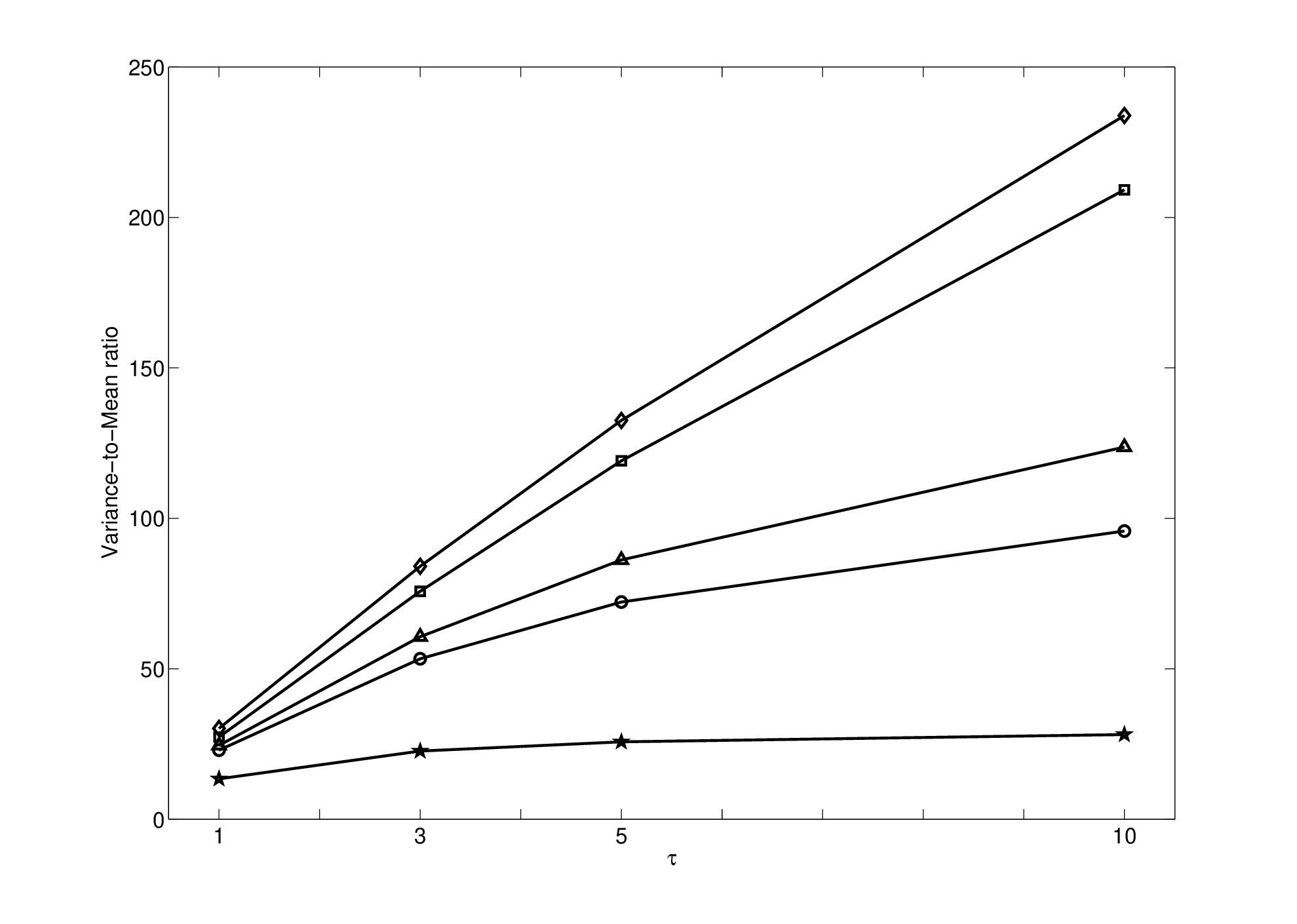}
\end{tabular}
\end{center}
\caption{\label{Fig:VtM2}Variance-to-Mean ratios at time instants $\tau=1,3,5,10$, for $4$ randomly selected $\map_2$ representations.}
\end{figure}

It is also known that the counting process in a general $m$-state $\map$ has dependent increments. In particular, for the $\map_2$ the covariance of the counting process between intervals of length $\tau$ is given by 
$$
C(\tau)=\pinegrita\left(M_1(\tau)\right)^2\enegrita-(\pinegrita M_1(\tau)\enegrita)(\pinegrita e^{D\tau}M_1(\tau)\enegrita),
$$
where $M_1(\tau)$ represents the first moment matrix of the counts during an interval $(0, \tau]$. Asymptotically, \cite{narayana} proved that 
$$
M_1(\tau) = E\left[N(\tau)\right]\enegrita \pinegrita+(\enegrita \pinegrita -D)^{-1}D_1 \enegrita \pinegrita + \enegrita (\pinegrita D_1(\enegrita \pinegrita-D)^{-1})-2\frac{E\left[N(\tau)\right]}{\tau}\enegrita \pinegrita.
$$
A measure of the potential overdispersion of the loss counting process is the Variance-to-Mean ratio \citep{lindsey}, defined as
\begin{equation}\label{vartomean}
VtM\left(N(\tau)\right)=\frac{V\left[ N(\tau)\right] }{E\left[N(\tau)\right]}.
\end{equation}
\cite{Nasr} recently define the Index of Dispersion of Counts (IDC) in the same way as in (\ref{vartomean}). In order to empirically study the range of values of (\ref{vartomean}), a total of $10000$ random set of parameters $\{x,y,u,v\}$ as in (\ref{can1})-(\ref{can2}) characterizing a $\map_2$ were simulated. Then, the Variance-to-Mean ratios as in (\ref{vartomean}) were calculated for an assortment of $\tau$ values ($\tau \in \{1,3,5,10\}$). Figure \ref{Fig:VtM} depicts the obtained results.  From the figure it is clear that there are multiple configurations of parameters $\{x,y,u,v\}$ under which $V\left[ N(\tau)\right]$ is considerably higher than $E\left[ N(\tau)\right]$, a phenomenon that intensifies as $\tau$ increases. This is better observed in Figure \ref{Fig:VtM2}, which shows the evolution of the Variance-to-Mean ratio with the value of $\tau$, for four different representations of $\map_2$s. It is interesting to note that in the simulation, only a $6.74\%$ of $\map_2$s led to a Variance-to-Mean ratio less than $1$. 
These preliminar results point out the suitability of the $\map_2$ for modeling overdisperse data. Further theoretical research needs to be carried out in this line but since it is not the scope of the paper, we leave it as an open task. 

\subsection{Persistence measures}
Next we shall study some persistence properties concerning the loss counting process of the $\map_2$. There are works in the literature which consider the modeling of persistence since this is an inherent phenomenon in many financial and insurance contexts, see for example \cite{vallois2009claims,stutzer2020persistence} and the references therein. In general, in the OpRisk (and any financial) context it might be of interest to infer the probability that the next loss occurs in a {\it short} (or {\it long)} time given that the previous loss was {\it recent} (or {\it far away}). Also, decision makers could benefit from the knowledge of estimates for consecutive long periods without losses (long inter-loss times spells) or in contrast, short inter-loss times spells which might result in a higher capital charge. We next define these event and provide formulae for their calculation.

Let $s$ denote a fixed time. Recall from Section 2 that $T_n$ is the time elapsing from the occurrence of the $(n-1)$-th loss until the next ($n$-th) loss. Define the following persistence measures as
\begin{eqnarray*}
p_{01}(s) &=& P\left(T_{n+1}>s \mid T_{n}<s\right),\\ \nonumber
p_{11}(s) &=& P\left(T_{n+1}>s \mid T_{n}>s\right).
\end{eqnarray*}
As previously commented, financial decision makers could be interested in estimating $1-p_{01}(s)$ for {\it small} values of $s$, or equivalently, the probability of the occurrence of a loss in a {\it short} period of time given that the previous loss occurred {\it recently}. Also, it would be advantageous to infer $p_{11}(s)$ for {\it large} values of $s$, that is, the probability of a {\it long} time up to the next loss given that the previous loss occurred {\it long ago}). In practice, $s$ can be chosen as a small or high percentile of a given historical trace of consecutive inter-loss times. Next result provides the expressions for the previous persistence measures.

\begin{proposition}\label{matrix_transition} Let a $\map_2$ be represented by $\{D_0,D_1\}$. Let $P^\star$ denote the transition matrix between states where losses occur and $\phinegrita$ its stationary distribution. Then,  
\begin{eqnarray}\label{p01}
p_{01}(s) &=& \frac{\phinegrita\left(I-e^{D_0 s}\right) P^\star e^{D_0 s}P^\star \enegrita}{1-\phinegrita e^{D_0 s}\enegrita},\\\label{p11}
p_{11}(s) &=& \frac{\phinegrita e^{D_0 s} P^\star e^{D_0 s}P^\star \enegrita}{\phinegrita e^{D_0 s}\enegrita},
\end{eqnarray}
for $s>0$.
\end{proposition}
See Appendix A for a proof.\\

Now, assume that the interest is to predict the probability of a sequence of {\it short (long)} inter-loss times spells.
As before, consider a fixed time $s$. Then, define two discrete random variables $S$ and $L$ (standing for {\it Short} and {\it Long}, respectively) as
\begin{eqnarray}\label{Shortspell}
S = n &\overset{d}{\equiv}& \left(T_1<s\right) \cap \left(T_2<s\right)\cap \ldots \cap \left(T_n<s\right) \cap \left(T_{n+1}>s\right),\\ \label{Longspell}
L = n &\overset{d}{\equiv}& \left(T_1>s\right) \cap \left(T_2>s\right)\cap \ldots \cap \left(T_n>s\right) \cap \left(T_{n+1}<s\right).
\end{eqnarray}
Defined as in (\ref{Shortspell})-(\ref{Longspell}), then for a {\it small} value of $s$, $S$ counts for the number of consecutive short inter-loss spells. On the other hand, if $s$ is {\it large} then, $L$ represents the number of consecutive long inter-loss spells. The following result expresses the probabilities of (\ref{Shortspell}) and (\ref{Longspell}) in matrix form. 

\begin{proposition}\label{matrix_spells} Let a $\map_2$ be represented by $\{D_0,D_1\}$. Let $P^\star$ denote the transition matrix between states where losses occur and $\phinegrita$ its stationary distribution. Then, for a fixed value  $s>0$ 
\begin{equation}\label{Ps}
P(S = n)=
\left\{
\begin{array}{lll}
\phinegrita \left[\left( I-e^{D_0s}    \right)P^\star  \right]^n e^{D_0 s} P^\star \enegrita, & \text{ if } & n\geq 1 \\
&&\\
\phinegrita e^{D_0 s} \enegrita,& \text{ if } & n=0.
\end{array}
\right.
\end{equation}

\begin{equation}\label{Pl}
P(L = n)=
\left\{
\begin{array}{lll}
\phinegrita \left( e^{D_0s}  P^\star  \right)^n\left( I-e^{D_0s}    \right)P^\star \enegrita, & \text{ if } & n\geq 1 \\
&&\\
1-\phinegrita e^{D_0 s} \enegrita,& \text{ if } & n=0.
\end{array}
\right.
\end{equation}

\end{proposition}
See Appendix B for a proof.

A remark is made at this point.
\begin{remark}
Propositions 1 and 2 are also true for higher-order $\map$s. This is because in the matrix expressions for the different persistence measures only $D_0$ is involved (and not $D_1$, or $D_k$ for $k\geq 2$ for higher-order $\map$s). Also, the definitions of $P^\star$ and $\phinegrita$ are common for $\map$s of all orders, see \cite{Chakravarthy}. In the case of higher-order $\map$s the proofs shall be identical but changing the counter of the summation from $i,j,k \in \{1,2 \}$ by $i,j,k \in \{1,2,\ldots,m \}$, being $m$ the order of the process.
\end{remark}

To conclude this section, we present several numerical results in order to illustrate some of the patterns that (\ref{p01})-(\ref{p11}) and (\ref{Ps})-(\ref{Pl}) may follow. This issue is of paramount importance to determine the versatility of the $\map_2$ for modeling real-life situations. In this respect, the presented results are very preliminar and further research should be carried out in future work.

In order to illustrate some of the patterns that (\ref{p01})-(\ref{p11}) and (\ref{Ps})-(\ref{Pl}) are able to model, four different $\map_2$ representations were chosen:

\begin{enumerate}
\item[R1.] 
\begin{equation*}
D_{0}=\begin{pmatrix}
  -1.1272 &\ 0.0055  \\
 0 &\ -42.4417 \\
\end{pmatrix},\quad
D_{1}=\begin{pmatrix}
 1.1217 &\ 0  \\
 0.2173 &\ 42.2244 \\
\end{pmatrix}.
\end{equation*}
\item[R2.] \begin{equation*}
D_{0}=\begin{pmatrix}
  -1.4373 &\ 0.0498  \\
 0 &\ -14.2706 \\
\end{pmatrix},\quad
D_{1}=\begin{pmatrix}
 1.3875 &\ 0  \\
0.5283 &\ 13.7423 \\
\end{pmatrix}.
\end{equation*}
\item[R3.]

\begin{equation*}
D_{0}=\begin{pmatrix}
  -0.6830 &\ 0.0026  \\
 0 &\ -34.6904 \\
\end{pmatrix},\quad
D_{1}=\begin{pmatrix}
 0 &\ 0.6804  \\
 34.5586 &\ 0.1318 \\
\end{pmatrix}.
\end{equation*}
\item[R4.]
\begin{equation*}
D_{0}=\begin{pmatrix}
  -0.9751 &\ 0.5933  \\
 0 &\ -46.6547 \\
\end{pmatrix},\quad
D_{1}=\begin{pmatrix}
 0 &\ 0.3818  \\
 10.7740 &\ 35.8806 \\
\end{pmatrix}.
\end{equation*}
\end{enumerate}
The previous selection of parameters was made among a total of one million simulated $\map_2$s so that the four sets of parameters lead to different forms for expressions (\ref{p01}), (\ref{p11}), (\ref{Ps}) and (\ref{Pl}).

\begin{figure}[h]
\begin{center}\begin{tabular}{cc}
\includegraphics[height=2.2in]{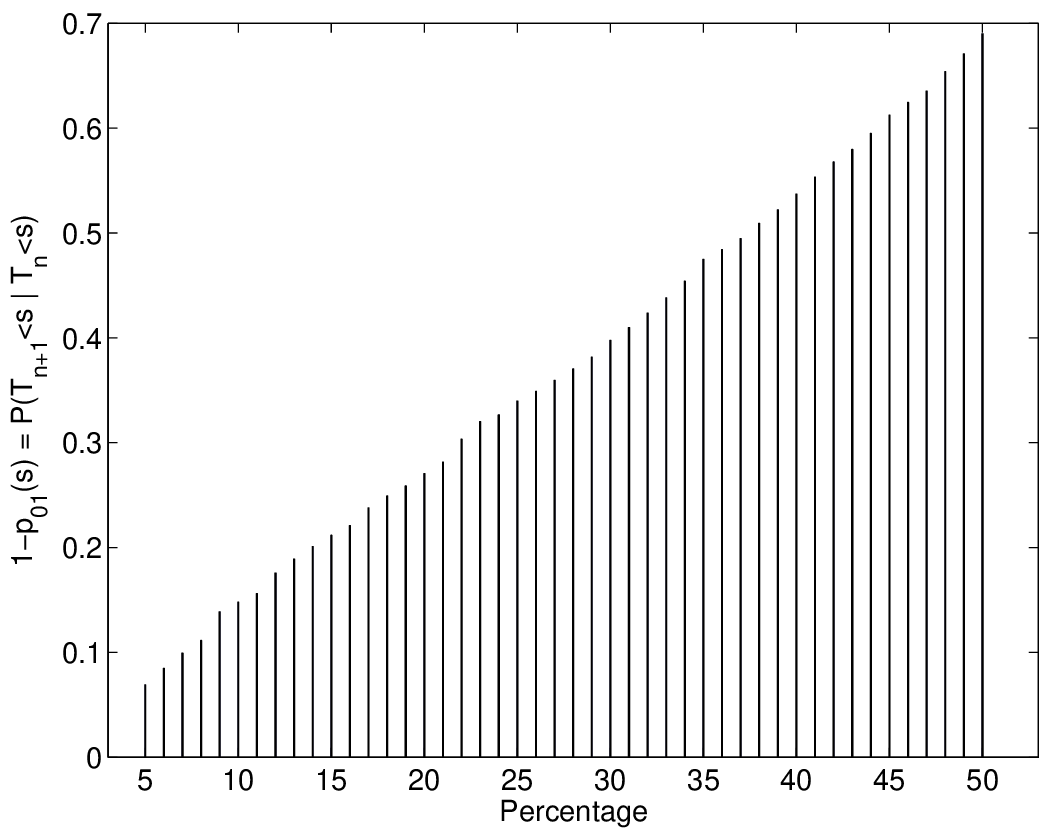}&\includegraphics[height=2.2in]{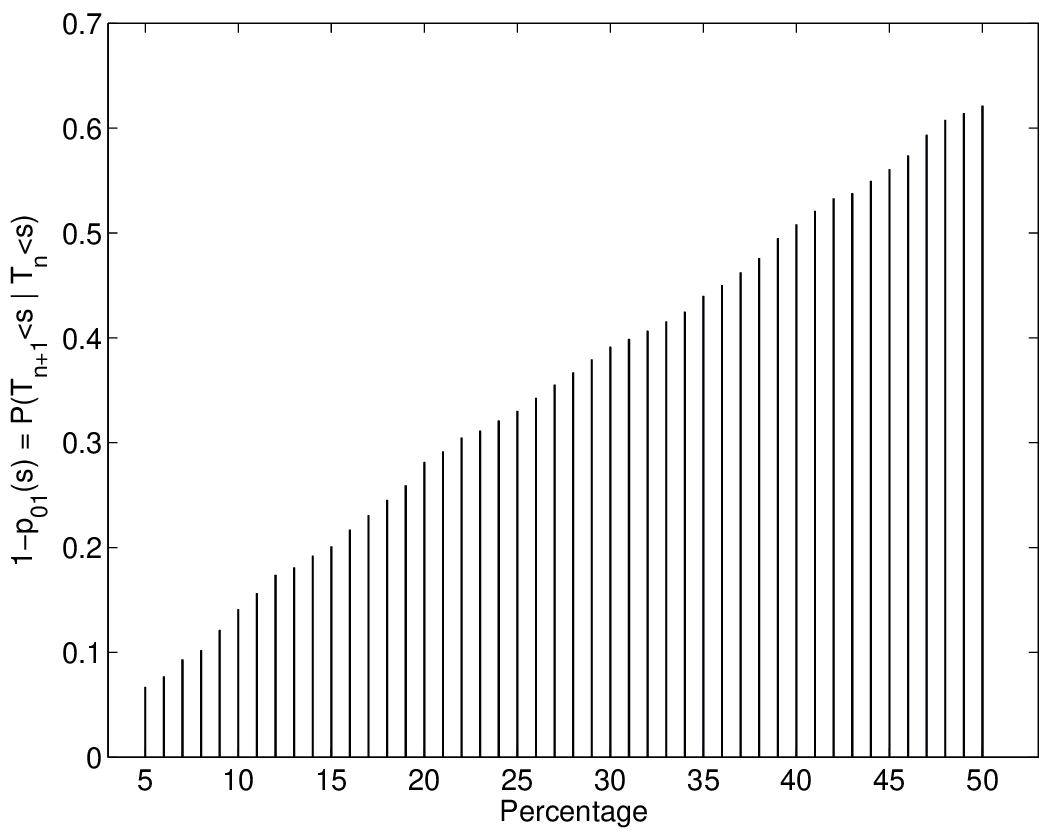}\\
\includegraphics[height=2.2in]{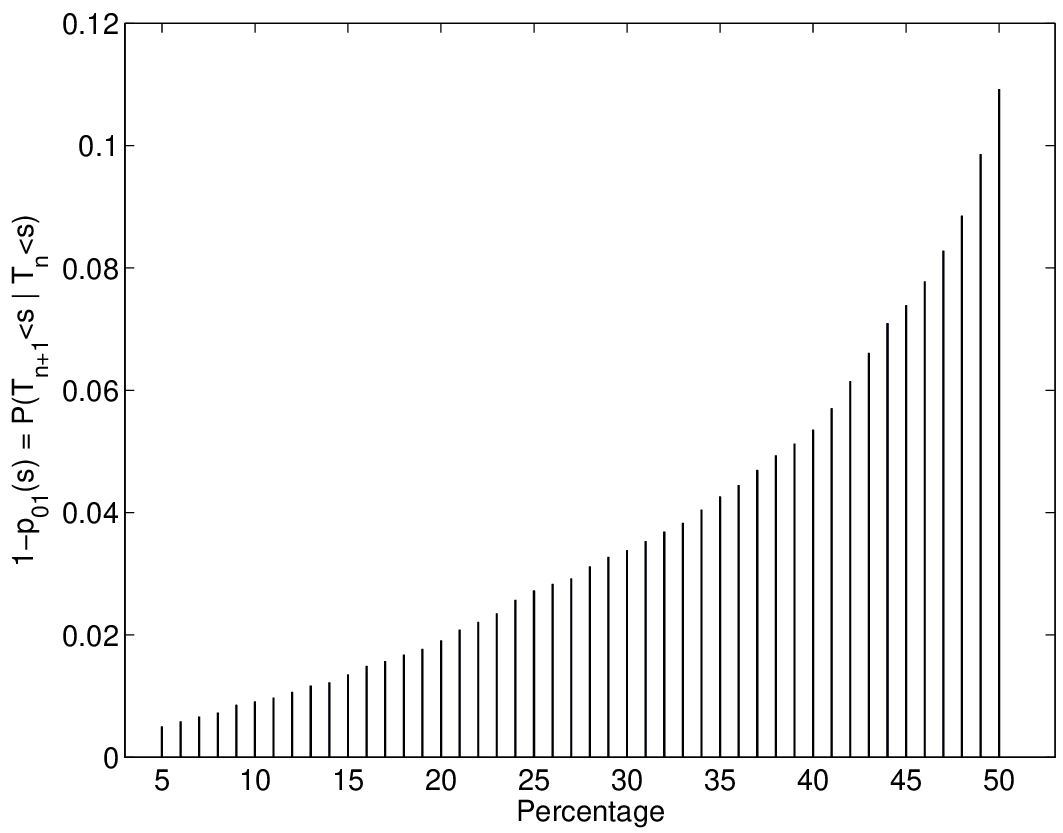}&\includegraphics[height=2.2in]{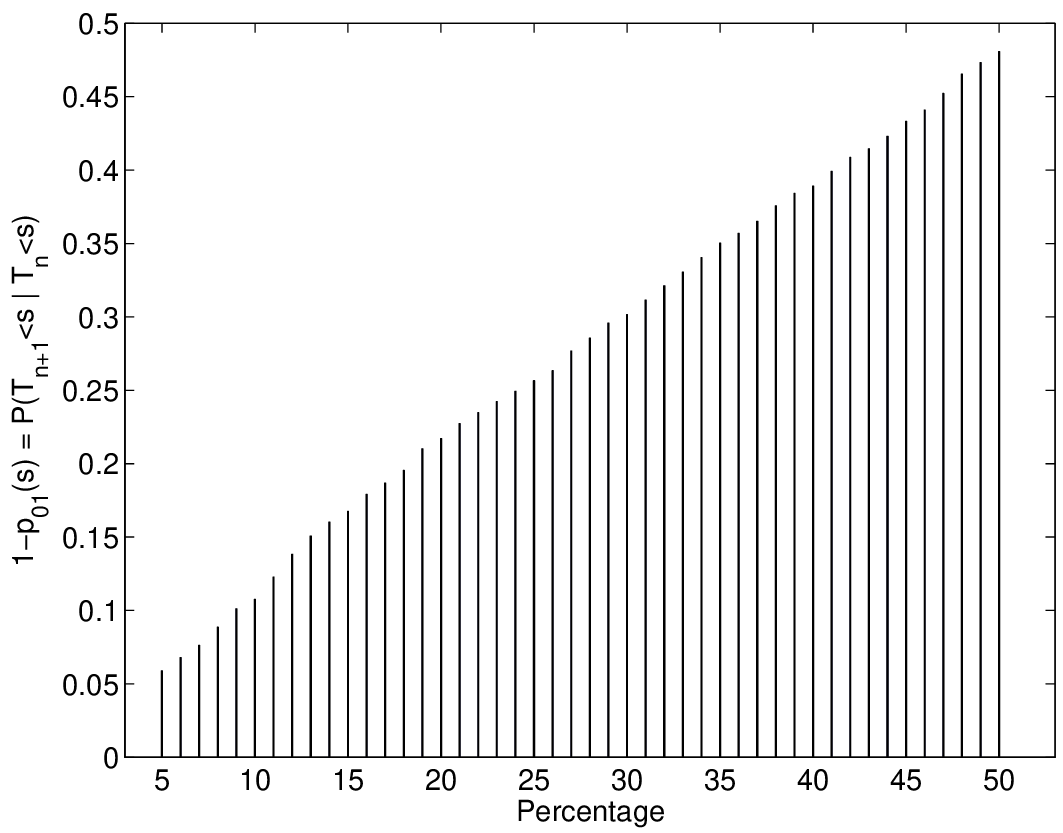}\\
\end{tabular}
\end{center}
\caption{\label{Figp01} $P(T_{n+1}<s\mid T_{n}<s)$ for the four $\map_2$ representations: R1 (top left panel), R2 (top right panel), R3 (bottom left panel) and R4 (bottom right panel). The value of $s$ is determined by the percentiles of percentages given by the $X$-axis.}
\end{figure}

\begin{figure}[h]
\begin{center}\begin{tabular}{cc}
\includegraphics[height=2.2in]{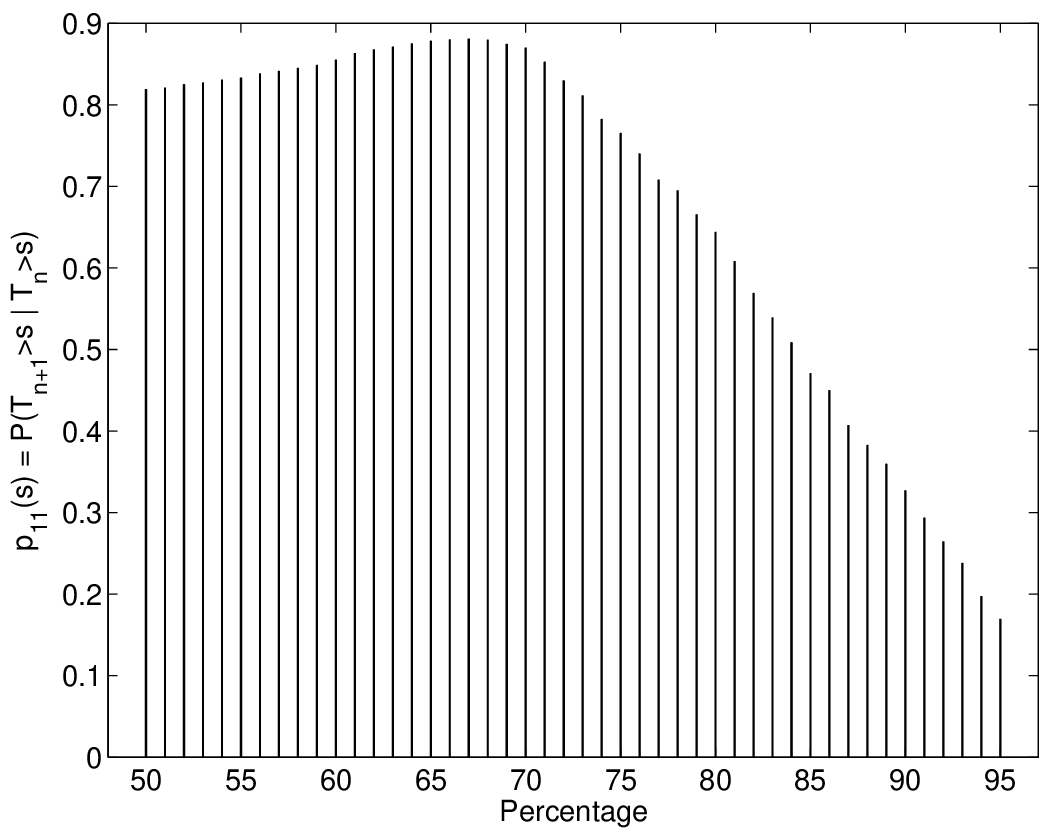}&\includegraphics[height=2.2in]{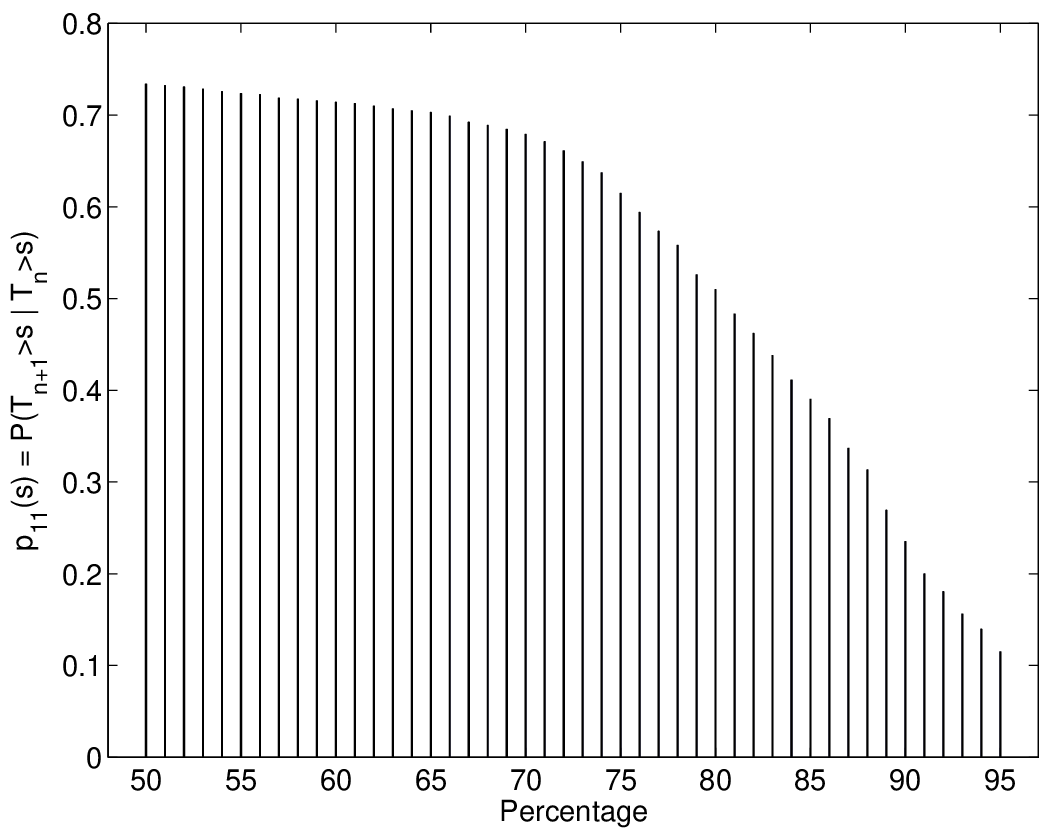}\\
\includegraphics[height=2.2in]{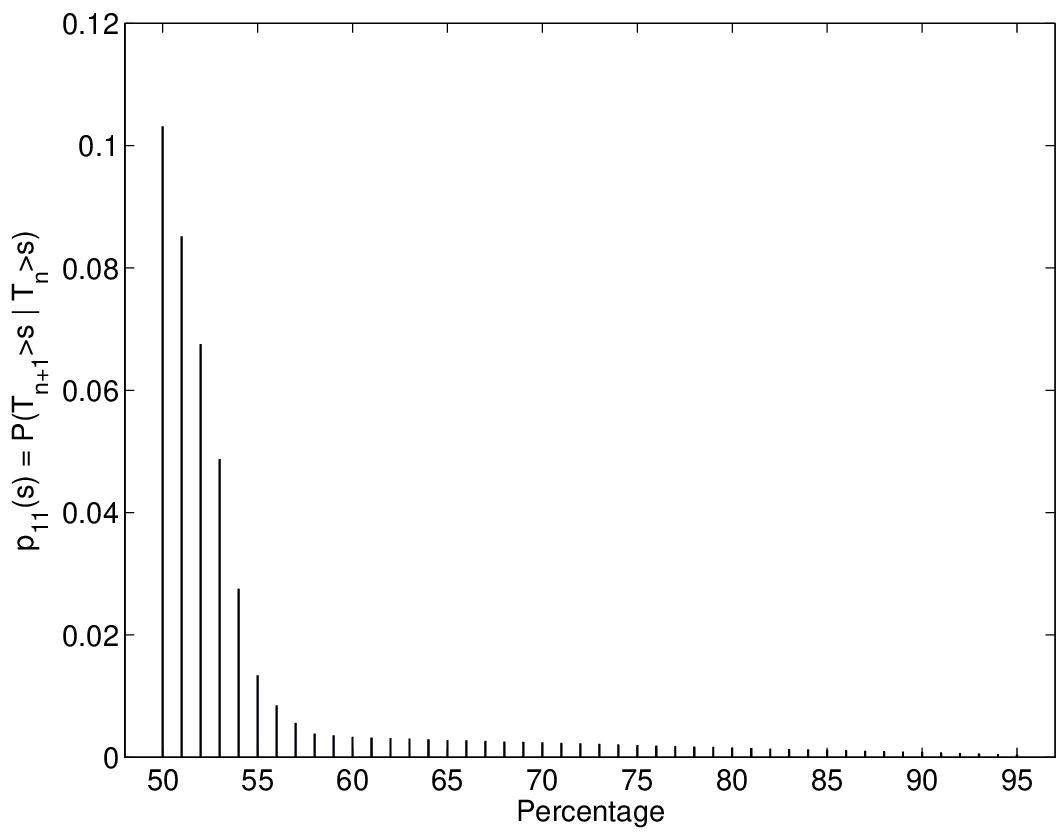}&\includegraphics[height=2.2in]{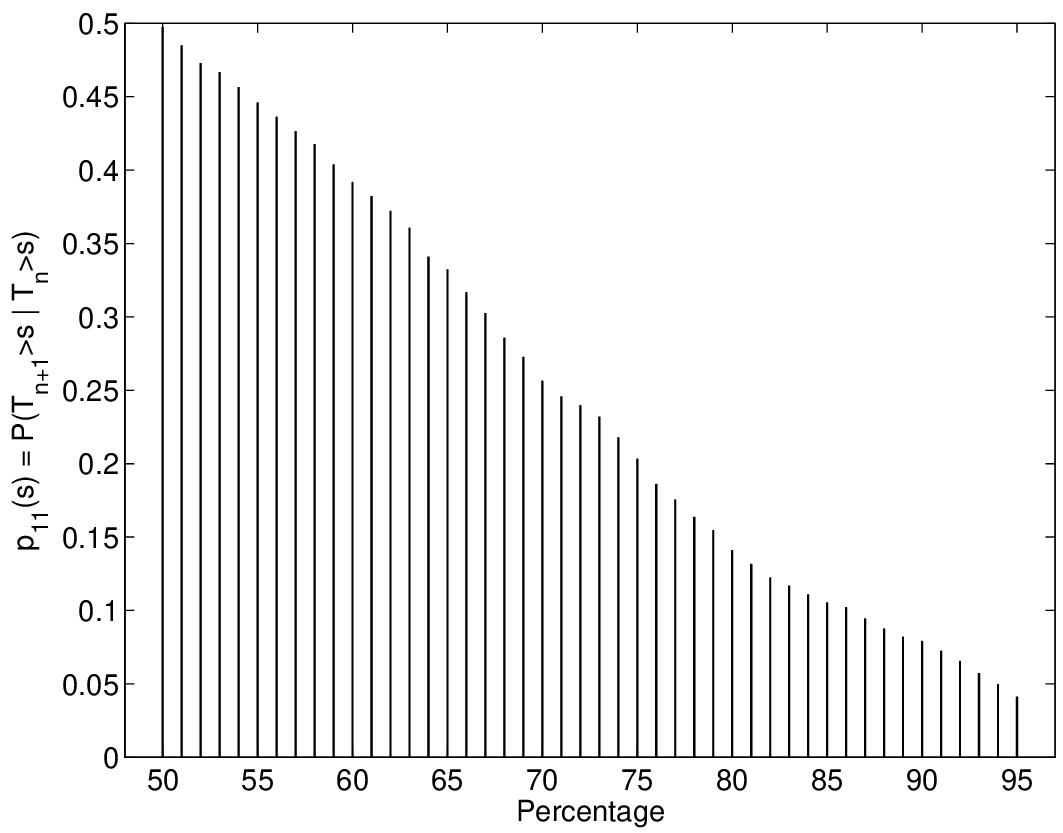}\\
\end{tabular}
\end{center}
\caption{\label{Figp11} $P(T_{n+1}>s\mid T_{n}>s)$ for the four $\map_2$ representations: R1 (top left panel), R2 (top right panel), R3 (bottom left panel) and R4 (bottom right panel). The value of $s$ is determined by the percentiles of percentages given by the $X$-axis.}
\end{figure}

For each representation, $1-p_{01}(s)$ and $p_{11}(s)$ are computed for values of $s$ equal to the low percentiles ($5$-th,$6$-th,...,$49$-th and $50$-th) in the first case, and high percentiles ($50$-th,$51$-th,...,$94$-th and $95$-th) in the second case. The results are shown in Figures \ref{Figp01} and \ref{Figp11}. Different models are obtained: in all the considered cases, the values of  $1-p_{01}(s)$ increase with $s$ although the increase pattern differs between representations. On the contrary, there are cases where $p_{11}(s)$ is not monotonic.

\begin{figure}[h]
\begin{center}\begin{tabular}{cc}
\includegraphics[height=2.2in]{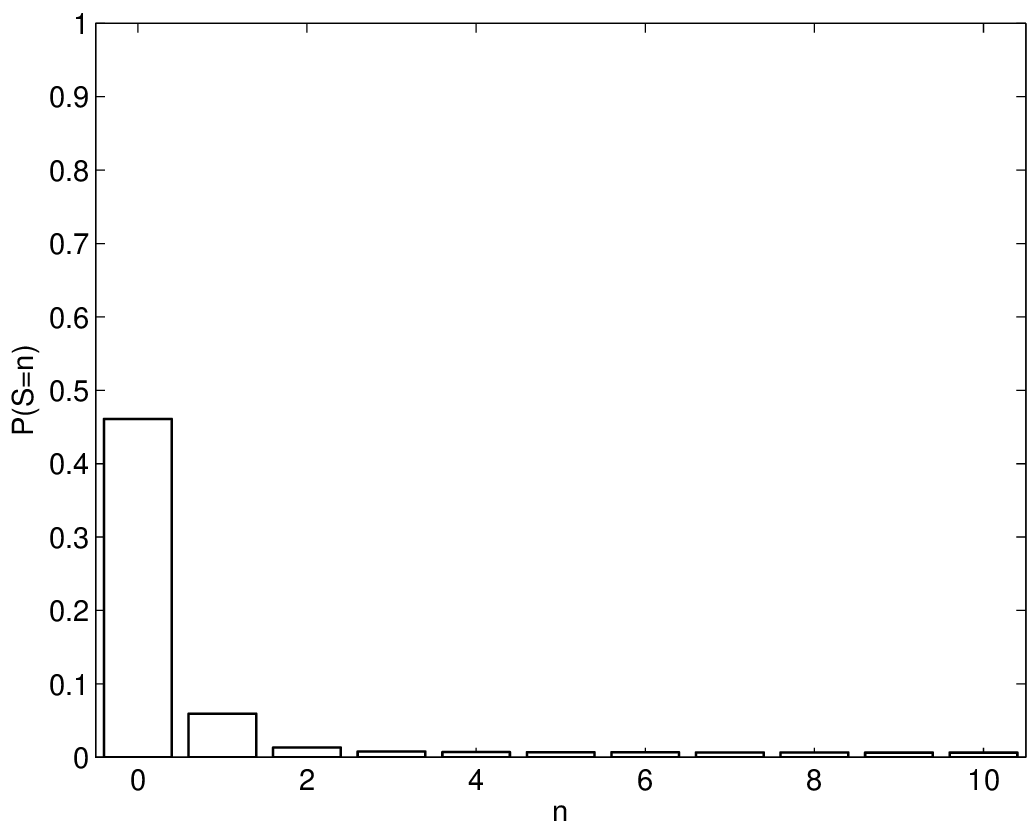}&\includegraphics[height=2.2in]{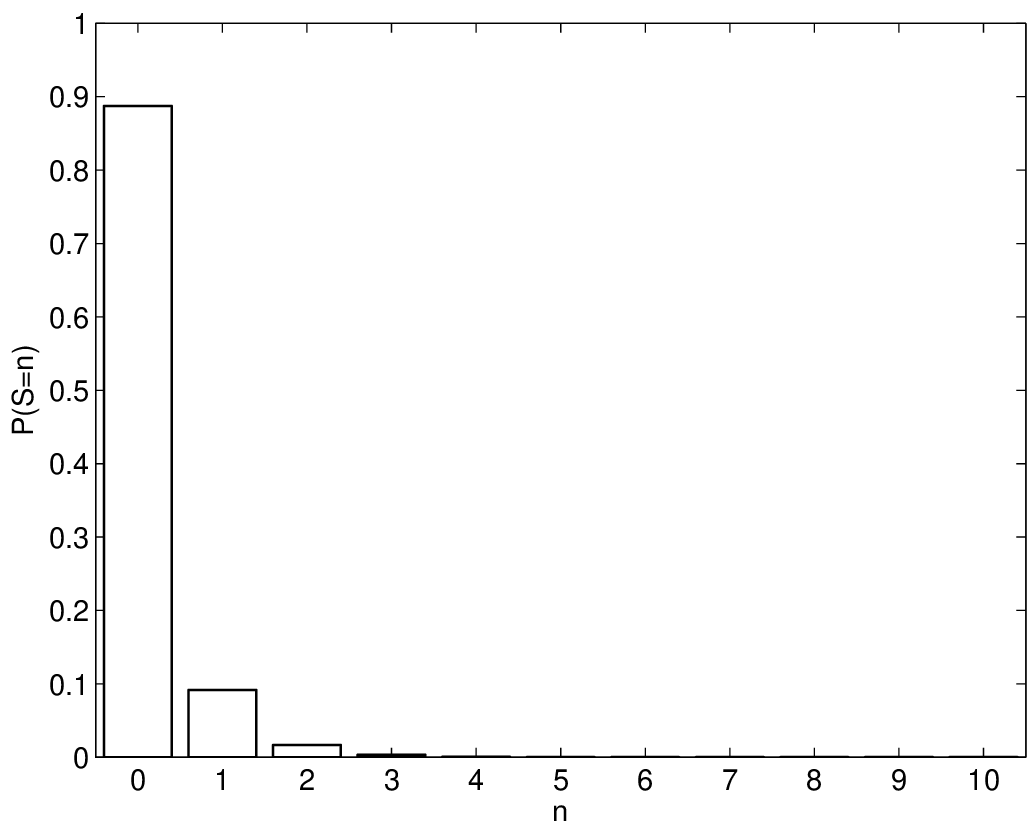}\\
\includegraphics[height=2.2in]{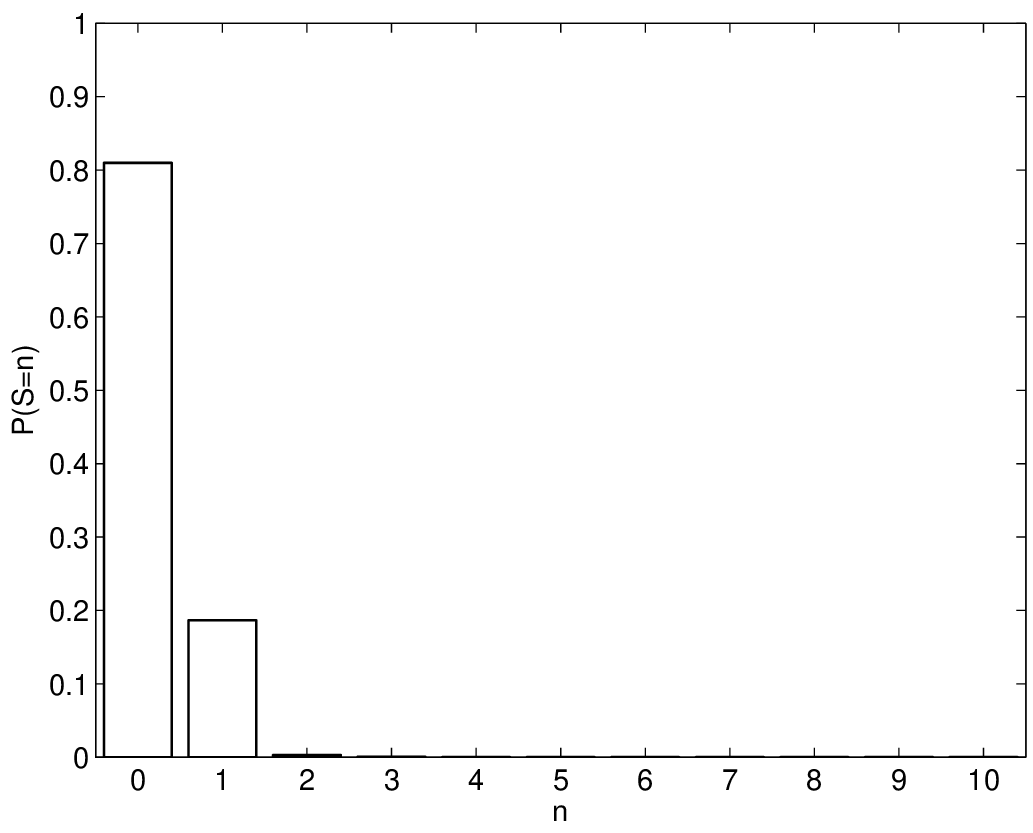}&\includegraphics[height=2.2in]{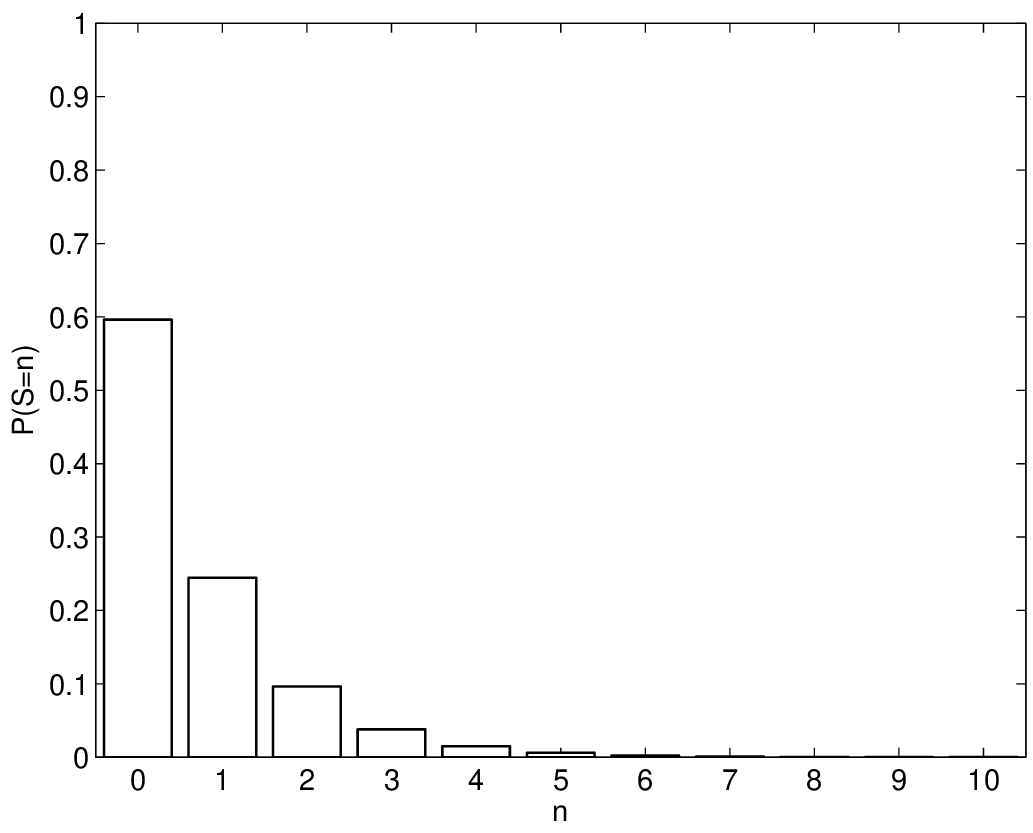}\\
\end{tabular}
\end{center}
\caption{\label{FigS}$P(S=n)$ for the four $\map_2$ representations: R1 (top left panel), R2 (top right panel), R3 (bottom left panel) and R4 (bottom right panel). }
\end{figure}

\begin{figure}[h]
\begin{center}\begin{tabular}{cc}
\includegraphics[height=2.2in]{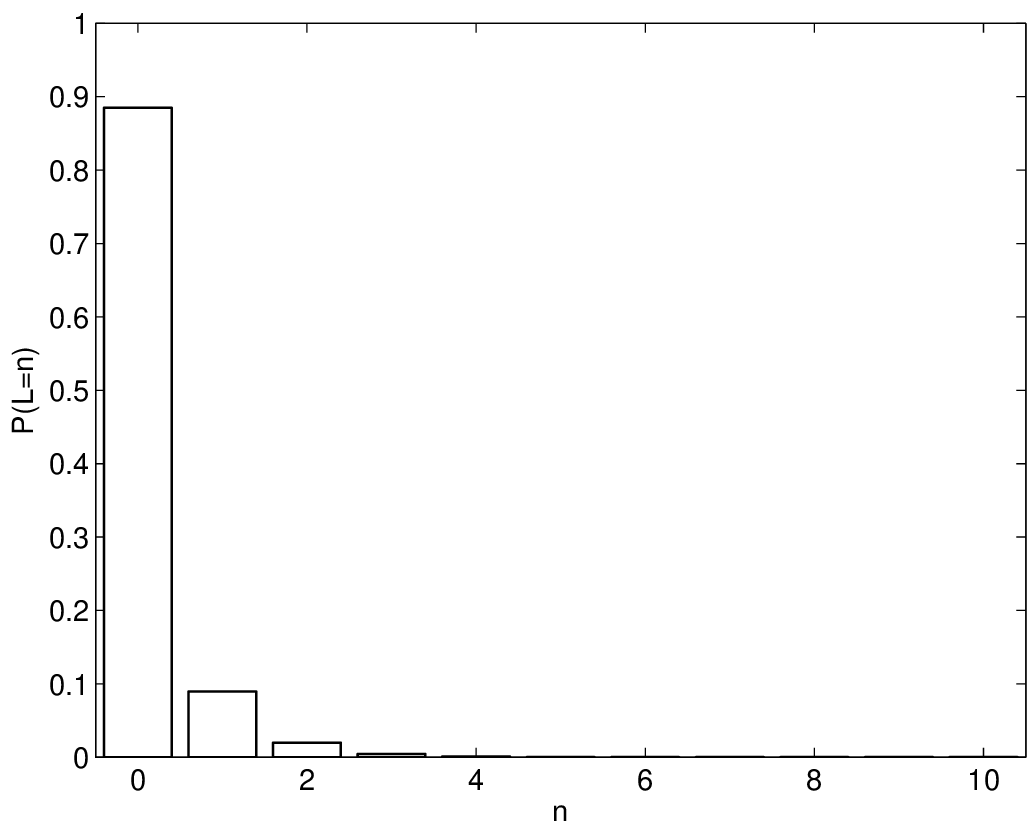}&\includegraphics[height=2.2in]{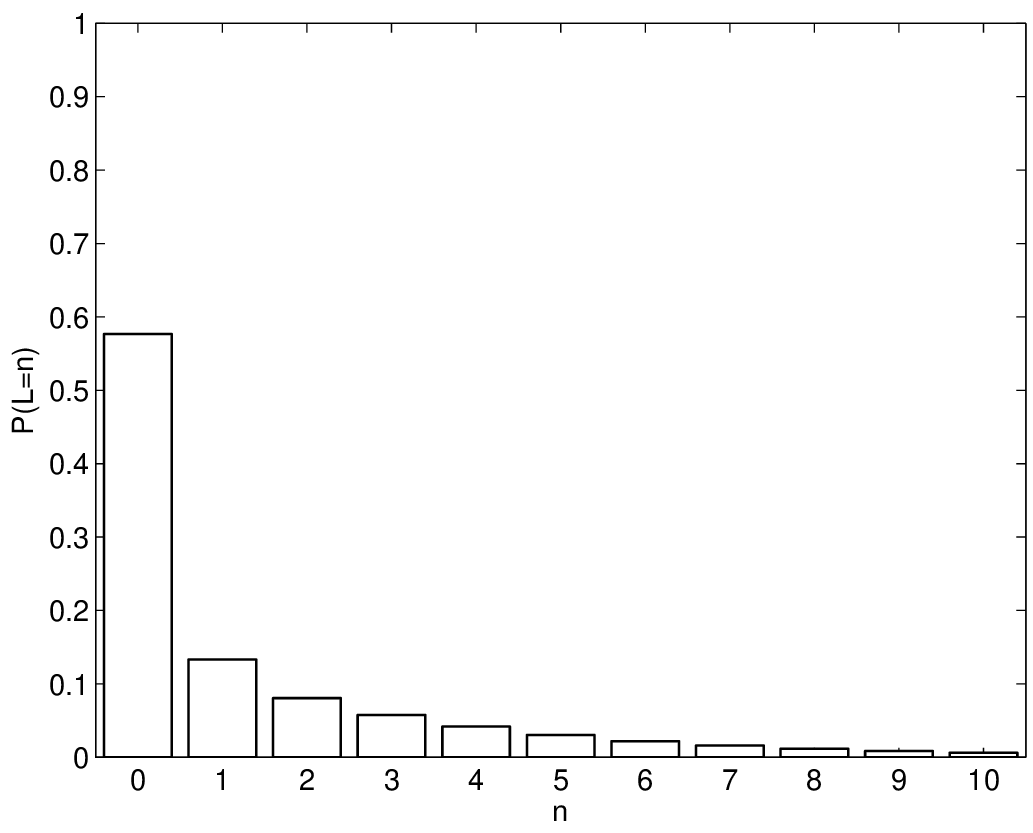}\\
\includegraphics[height=2.2in]{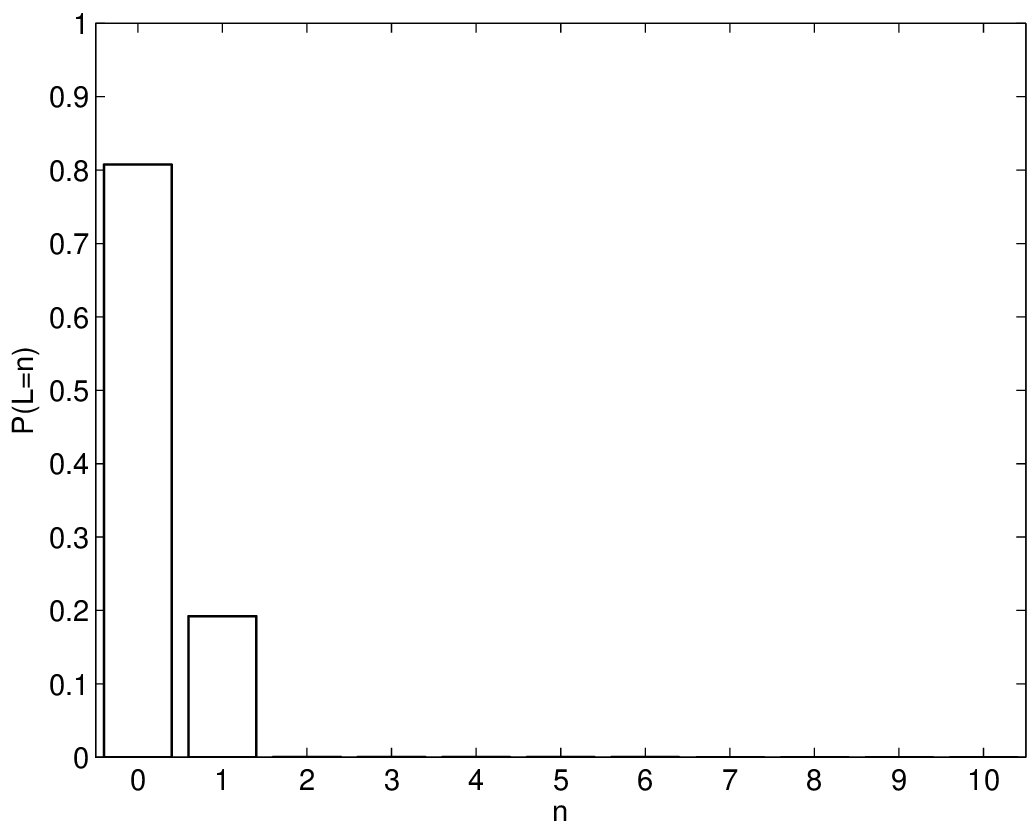}&\includegraphics[height=2.2in]{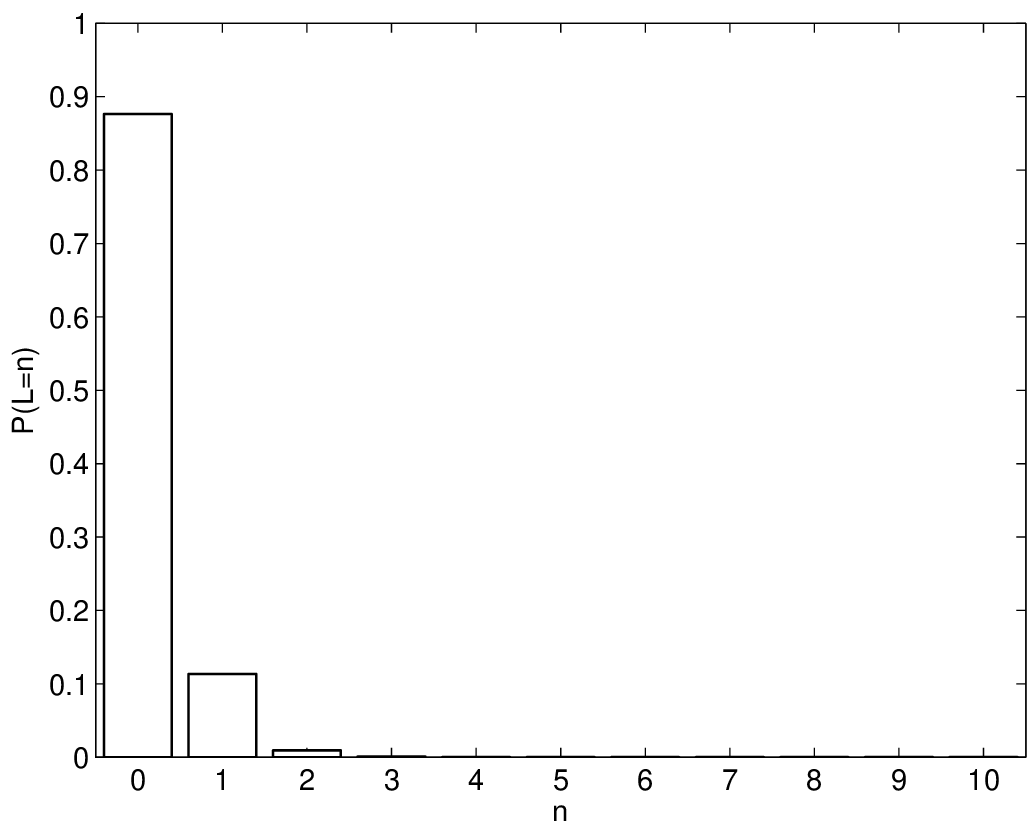}\\
\end{tabular}
\end{center}
\caption{\label{FigL}$P(L=n)$ for the four $\map_2$ representations: R1 (top left panel), R2 (top right panel), R3 (bottom left panel) and R4 (bottom right panel).}
\end{figure}

Figures \ref{FigS} and \ref{FigL} depict the probabilities of the random variables that count for the number of {\it short} and {\it long} inter-loss times. In particular, the value of $s$ has been chosen as the $20$-th and $80$-th percentiles of the inter-loss times distributions, respectively for $S$ and $L$. In the four cases, the probabilities mass functions are found to be unimodal at $0$ and rapidly decrease towards $0$, a pattern that was found in the total of the one million simulated $\map_2$s.

\section{{Modeling the OpRisk}}\label{review_loss}
According to the European Union Solvency II Directive for insurers \citep{anexo_european}, the OpRisk is defined as ``\emph{the risk of a change in value caused by the fact that actual losses, incurred for inadequate or failed internal processes, people and systems, or from external events, differ from the expected losses}''. Because of its high expected impact to the banking industry, since the 9/11 attacks and tradings at Soci\'et\'e G\'en\'eral, AIB and National Australia Bank, researchers have been studying approaches for properly modeling the OpRisk. The Basel II framework forces financial institutions to set aside a capital charge against unexpected losses derived from OpRisk. Three different methodologies have been proposed in the literature for OpRisk quantification: the Basic Indicator Approach, the Alternative Standardized Approach and, the most sophisticated one (in terms of statistical modeling), the Advanced Measurement Approach (AMA), where the Loss Distribution Approach (LDA) is included in. 
In the LDA the estimation of the annual loss distribution $Z$ as in (\ref{risk_model}) is achieved by modeling separately the distributions of the number of losses per year ($N$) and the severities ($X_i$), under the assumption of independence between them.
It is common that the Poisson process or the Negative Binomial distribution are chosen to model the frequency, while some type of heavy-tailed distribution is selected to fit the severities, \cite{Chernobai1,bolance2012quantitative}. For a thorough treatment  of the OpRisk modeling, we refer the reader to \cite{panjer2006operational,Shevchenko1,bolance2012quantitative,peters2015advances,Bimaquest}.


The cumulative distribution function of $Z$ is calculated as
$$
F_Z(x) = P\left(Z\leq x  \right) =\sum_{n=0}^{\infty}P(N=n) F_n^{\star}(x),
$$
where $F_n^{\star}(x) = P\left(X_1+\ldots,X_n\leq x\right)$ is the $n-$th convolution of the severities' distribution. Although a closed-form expression for the loss aggregate distribution is usually hard to be obtained, its first and second moments can be computed in terms of those of the severity and frequency distributions as
\begin{eqnarray*}
E(Z)& = &E(N)E(X_1),\\
V(Z) &=& E(N)V(X_1)+V(N)E(X_1)^2.
\end{eqnarray*}
Closed-form expressions for higher order moments can be found, for example in \cite{{Shevchenko2}}. Under (\ref{risk_model}), the capital charge is obtained as the Value-at-Risk (VaR), 
\begin{equation}\label{var}
VaR_{p}\left(Z\right) = F_{Z}^{-1}(p),
\end{equation}
where $F_{Z}^{-1}$ is the inverse cumulative distribution function of the total annual loss, and $p$ denotes the risk tolerance which is usually chosen close to 1. Another measure of interest in the OpRisk modeling is the expected shortfall (equivalently, conditional VaR), which unlike the VaR, satisfies the property of coherence. It is computed as
\begin{equation}\label{es}
E\left(Z\mid Z\geq VaR_p(Z)  \right).
\end{equation}
If the distribution of $Z$ has an infinite mean, then clearly the expected shortfall does not exist. 

In order to approximate (\ref{risk_model}) a battery of numerical methods, either based on simulations, inversion of characteristic functions, or recursions, have been proposed in the literature. For an exhaustive review of such methods, see \cite{panjer2006operational,Shevchenko2,Shevchenko1,Gerhold}, or Ch. 9 in \cite{klugman2012loss}.

%
\section{Numerical application}\label{numerical}

{In this section we illustrate an application of the $\map_{2}$ in the context of the OpRisk. In particular, given a real OpRisk database, described in Section \ref{dataset}, the $\map_2$ is fitted to the times between consecutive losses in Section \ref{est_freq}. Descriptors of interest associated to the distributions of the inter-loss times and the number of yearly losses, as well as estimations for the persistence measures described in Section 3 are obtained. In Section \ref{est_sev} the severities' distribution are modeled through the heavy-tailed, double Pareto Lognormal distribution. As will be seen, the combination of the $\map_2$ for the frequency with the heavy-tailed model for the loss severity distribution, will enable us to give an estimate of the loss aggregate model and the capital charge in Section \ref{est_compound}.}


\subsection{Data description}\label{dataset}
\begin{figure}[!h]
\begin{tabular}{c}\hspace{-1.5cm}
\includegraphics[height=2.6in]{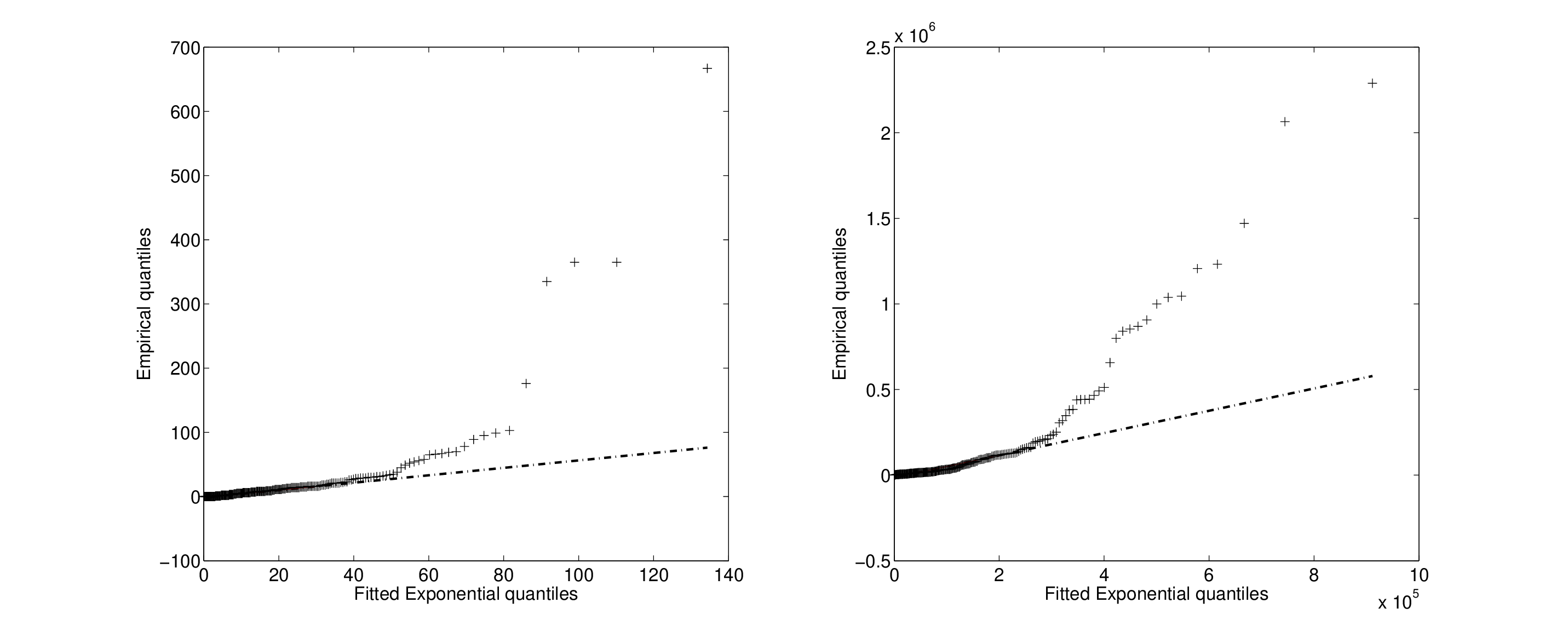} 
\end{tabular}
\caption{\label{Fig2} QQ-plots of the empirical inter-loss times (\emph{left panel}) and severities (\emph{right panel}) quantiles versus those of the fitted exponential distributions.}
\end{figure}

The real OpRisk data set considered in this paper consists of $225$ records of loss occurrences and associated severities in a Spanish financial institution. The sample, which corresponds to a unique risk cell (retail banking Basel II business line), was obtained from $4929$ consecutive days  ($13.5$ years) in the period between 30/12/1993 and 29/06/2007. In a total of $225$ out of the $4929$ days, a single loss occurred which was recorded together with the corresponding severity. Some descriptors of the dataset are as follows. The average, median, variation coefficient and maximum value of the inter-loss times are $22.0045$, $8$, $2.827$ and $667$ days. Also, the empirical correlation coefficient is non-negligible and given by $0.3546$. For the severities, the average, median, variation coefficient and maximum are $151710$, $34318$, $2.0973$ and $2289060$. Note that, from the empirical values both samples are right-skewed with a tail longer than that of an exponential distribution. This can also deduced from Figure \ref{Fig2}, which depicts the empirical quantiles in comparison to those of the fitted (via MLE) exponential distributions. Note how the larger empirical quantiles are far from the fitted ones.

\subsection{The estimated frequency distribution}\label{est_freq}
In this section we describe how the database inter-loss times are modeled by the $\map_2$. The estimated (joint) distribution of the times, the modeling of the counting process $N(\tau)$ and of the persistence measures described in Section 3 will be analyzed. 


\subsubsection*{The distribution of the inter-loss times}
The methodology from Section \ref{inference2} was applied to fit a $\map_{2}$ to the sample of inter-loss times, and the obtained results were as follows. First, the estimated process was found as
\begin{equation}\label{estimated_map}
\widehat{D}_{0}=\left(\begin{array}{cc} -0.0063 & 0.0011\\
0 & -0.1036\end{array}\right), \qquad \widehat{D}_{1}=\left(\begin{array}{cc} 0.0052 & 0  \\
0.0016 & 0.1020 \end{array}\right).
\end{equation}
Very closed values were obtained if instead the Bayesian algorithm by \cite{pepaBayes} is considered.

\begin{figure}[!h]
\begin{center}\begin{tabular}{c}
\includegraphics[height=2.6in]{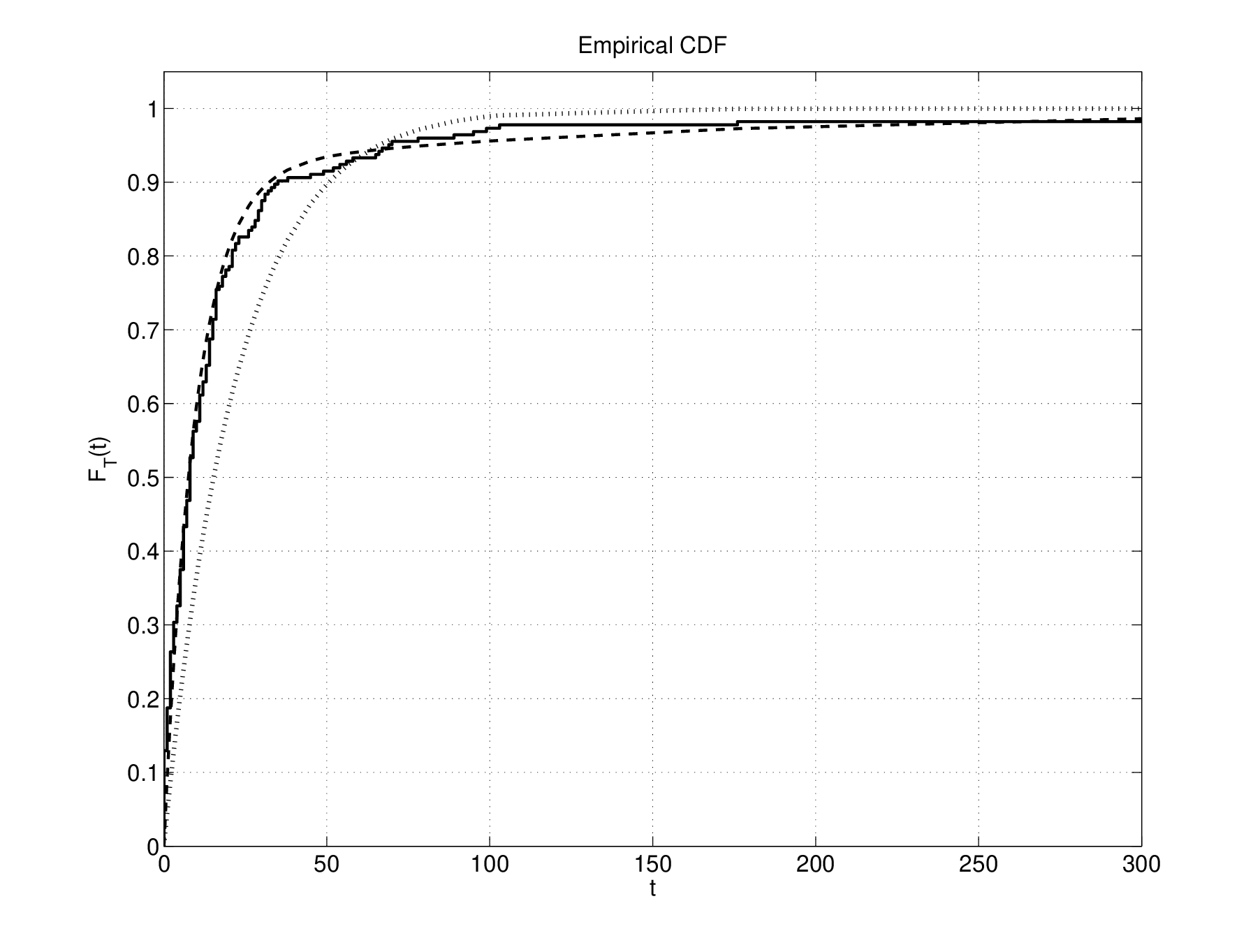} 
\end{tabular}
\end{center}
\caption{\label{Fig4} Inter-loss times cdf: empirical (solid line) versus estimated by the $\map_{2}$ (dashed line) and, by the exponential distribution (dotted line).}
\end{figure}

The mean, median, coefficient of variation and correlation coefficient under the estimated model were $22.0047$, $7.52$, $2.8205$ and $0.3545$, all close to the empirical values commented in Section \ref{dataset}. The estimated $\map_{2}$ (\ref{estimated_map}) defines a theorical cdf as in (\ref{cdf_ph}) which is plotted in dashed line in Figure \ref{Fig4} together with the empirical cdf of the data (solid line). The performance of the exponential distribution is also shown in dotted line. The outperformance of the PH-distribution (which defines the inter-loss times distribution in a $\map_{2}$) over the classic exponential distribution is clear from this figure.

\subsubsection*{The distribution of the number of losses per year}
We turn out to the estimation of the counting process. As commented in Section \ref{dataset} losses were recorded in a period of length between $13$  and $14$ years. It was found that the average number of losses per year is equal to $16.6154$, and the variance is $342.9231$. In this respect, the Poisson process presents again a poor performance, since it would notably underestimate the sample variance. Given the estimated $\map_2$ (\ref{estimated_map}), the expected number of losses as well as the variance were computed according to (\ref{expected_losses})-(\ref{variance_count}), and found as $16.5874$ and $240.0192$. 



\begin{figure}[h!]
\begin{center}\begin{tabular}{c}
\includegraphics[height=2.6 in]{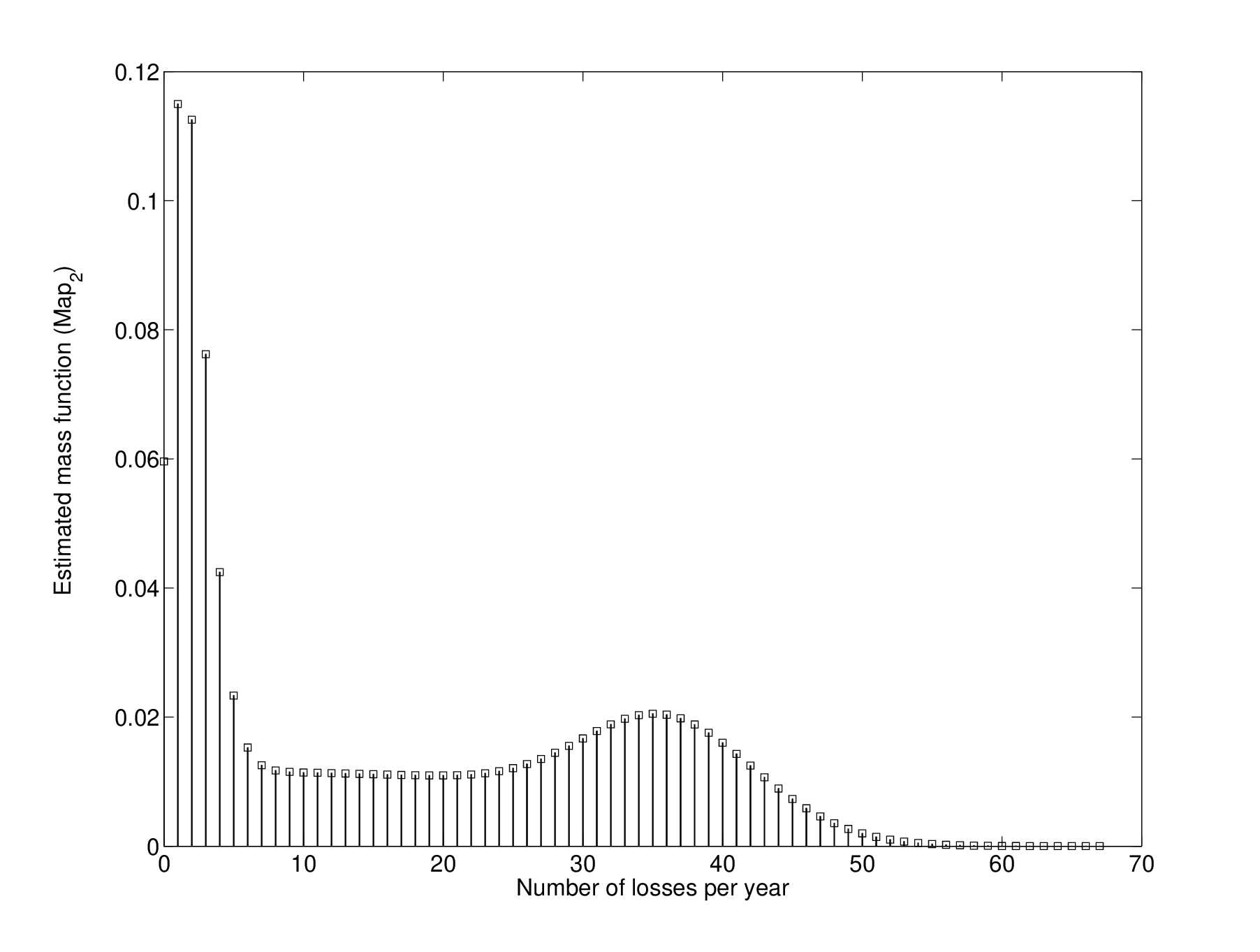}
\end{tabular}
\end{center}
\caption{\label{Fig:mass}Number of losses per year. Estimated mass function under the $\map_2$.}
\end{figure}

Since the modeling of the frequency in the OpRisk context usually refers to the one year horizon, the numerical algorithm described in \cite{joanna2} was applied to obtain the estimates of the probabilities as in (\ref{PN}), for $\tau=365$. {Hence, from now on, $N$ will denote $N(365)$}. Figure \ref{Fig:mass} depicts the obtained mass function of $N$, the annual number of losses, which will be used in Section  \ref{est_compound} for approximating the loss aggregate distribution. Note that the obtained probabilities cannot be compared against their empirical versions because of the small sample size (a sample of $13$ years). We finally would like to point out the differences between the adjusted mass function and that of a Poisson distribution. The first is unimodal around $0$, but it is non-negligible for high losses; in contrast, the Poisson mass function of rate $16.6154$ (equal to the average number of annual losses) concentrates around its mean with a much smaller variance.

\subsubsection*{Persistence measures}\label{persistence_measures_est}
Here we obtain the estimates of (\ref{p01})-(\ref{p11}) and (\ref{Ps})-(\ref{Pl}) for the real inter-loss times, according to the MLE estimates (\ref{estimated_map}).

\begin{figure}[!h]
\begin{center}\begin{tabular}{c}
\includegraphics[height=2.6in]{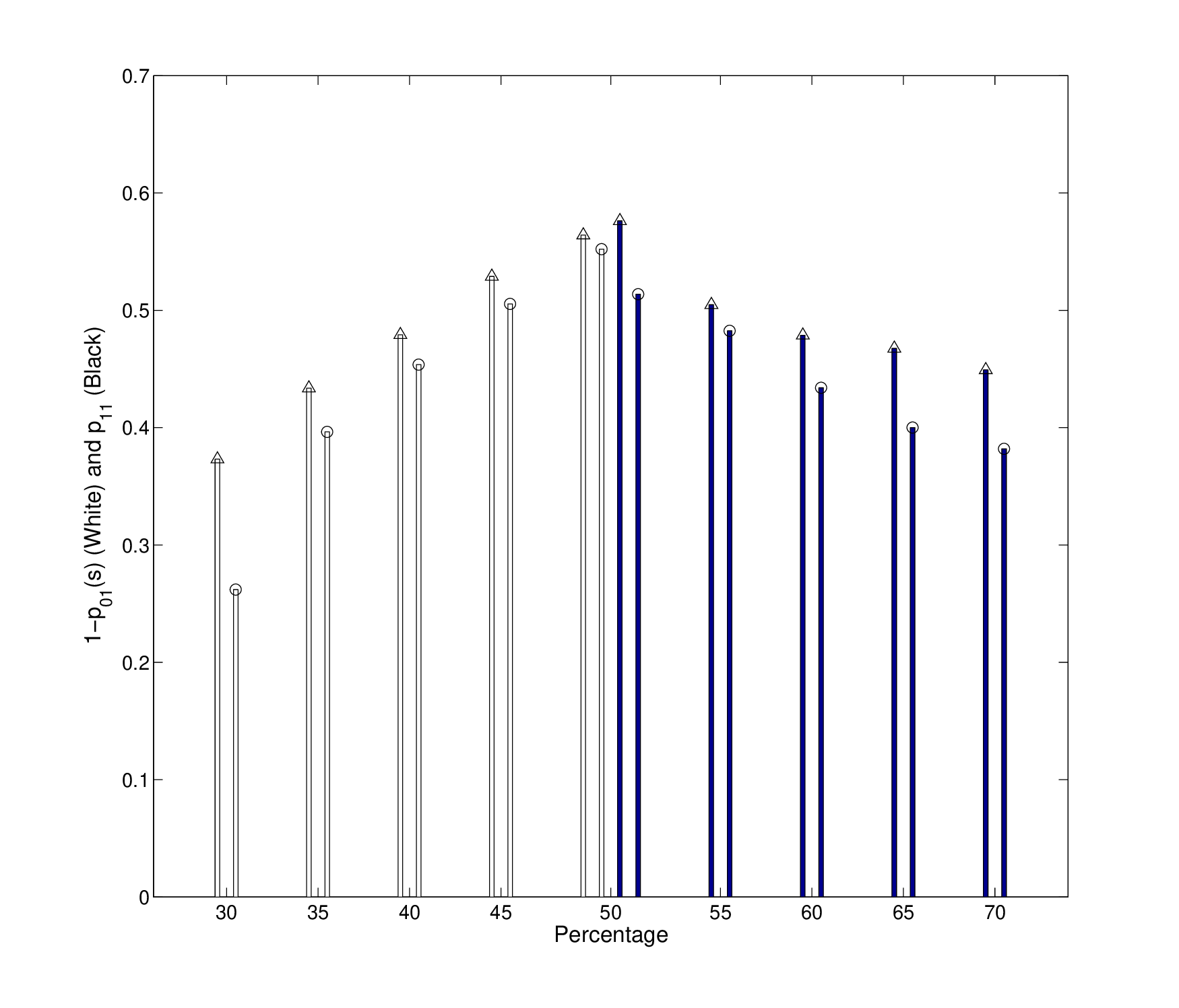} 
\end{tabular}
\end{center}
\caption{\label{Figtransition} White bars: $P(T_{n+1}<s\mid T_{n}<s)$, for $s=30,35,40,45,50$-th percentiles. Blues bars: $P(T_{n+1}>s\mid T_{n}>s)$, for $s=50,55,60,65,70$-th percentiles. Triangles denote empirical values and circles denote estimated values.}
\end{figure}

Consider first the estimation of the conditional probability of a next {\it short} ({\it long}) inter-loss time assuming that the previous inter-loss time was {\it short} ({\it long}). It is importante to remark that due to the small sample sizes, the empirical values of the transition probabilities $1-p_{01}(s)$ were not calculated for values of $s$ less than the $30$-th percentile of the inter-loss times distribution. In analogous way,  $p_{11}(s)$ was not estimated from the sample for values of $s$ larger than the times distribution's $70$-th percentile. As a result, the empirical values of $1-p_{01}(s)$ were estimated for $s$ equal to the $30$-th, $35$-th,...,$50$-th percentiles. These percentiles define for each case the length of the {\it short} inter-loss time. Similarly, the empirical probabilities $p_{11}(s)$ were estimated for $s$ ranging between the $50-$th, $55$-th,...,$70$-th percentiles (defining the {\it long} times). The empirical values (represented with a triangle) and the estimated values (represented by a circle) are depicted by Figure \ref{Figtransition}. In the figure, the white bars refer to the $1-p_{01}(s)$ values while the black bars concern the probabilities $p_{11}(s)$. Consider for example, the $30$-th percentile of the dataset (found as $s=3$). Then, the empirical value of $P(T_{n+1}<3\mid T_{n}<3)$ was $0.3731$ (that is, the $37.31\%$ of inter-loss times satisfying $T<3$ were followed by times that also satisfied the condition). This value is represented by the first white bar ending in a triangle. In this case, the fit given by the $\map_2$ (equal to $0.262$ and represented by the next white bar ending in a circle), clearly underestimated the empirical value. However, there are examples where the $\map_2$ provides good estimates. For instance, if $s$ is chosen as the $60$-th percentile (found as $s=11$), then the empirical probability $P(T_{n+1}>11\mid T_{n}>11)$ was $0.4787$, and the $\map_2$ provided an estimated value equal to $0.4340$. 

\begin{figure}[h!]
\begin{center}\begin{tabular}{cc}\hspace{-1cm}
\includegraphics[height=2.4in]{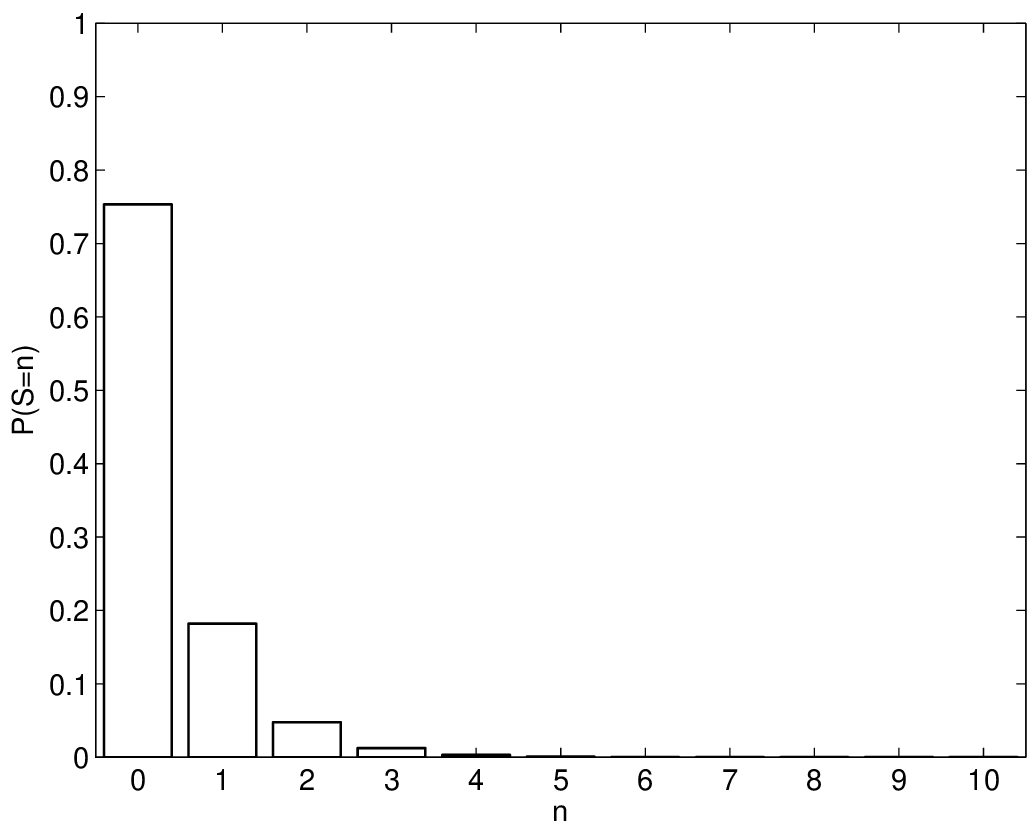}&\includegraphics[height=2.4in]{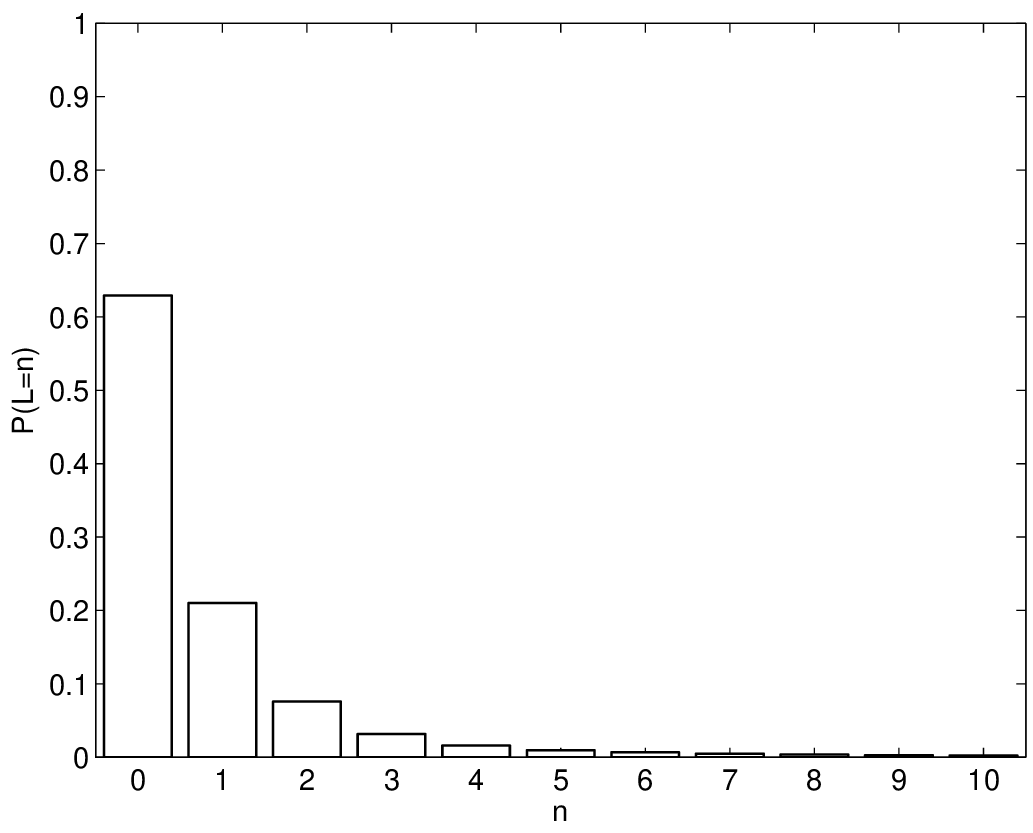}
\end{tabular}
\end{center}
\caption{\label{Figspells} Estimated values for the mass probability functions of the {\it short} ($S$, left panel) and  {\it long} ($L$, right panel) spells of the OpRisk dataset.}
\end{figure}

Consider next the estimation of the probabilities of short and long spells, as in (\ref{Ps})-(\ref{Pl}). As in the previous case, $s=3$ is considered for the study of $S$ (sequence of {\it short} spells), and $s=11$ for that of $L$ (sequence of {\it long} spells). Figure \ref{Figspells} depicts the estimated probabilities for values of $n=0,1,\ldots,10$. For both examples, the functions are unimodal at $0$ but with non-negligible values for the cases $n=1,2$. The empirical values for the probabilities $P(S=0)$, $P(S=1)$ and $P(S=2)$ were $0.7366$, $0.0804$ and $0.0670$. The estimated values, shown by the first three bars in the left panel of the figure, where $0.7535$, $0.1419$ and $0.0476$, respectively. If the interest is in the sequence of {\it long} spells, the empirical values for the probabilities $P(L=0)$, $P(L=1)$ and $P(L=2)$ were $0.6161$, $0.1875$ and $0.0580$ and the estimated ones were $0.6291$, $0.2101$ and $0.0759$.

\subsection{The estimated severities' distribution}\label{est_sev}
Concerning the severity distribution, we fit the data by a double Pareto Lognormal (\emph{dPlN}) distribution. Such distribution has been suggested in the literature as a versatile heavy-tailed model able to correctly model both the tail and body of the distribution and also, to capture different forms of asymmetry. See for example, \cite{Reed} or \cite{PepadPlN}. We briefly recall its definition. A random variable $X$ follows a double Pareto Lognormal distribution with parameters $\left(\alpha,\beta,\mu,\sigma^2\right)$ if $X=e^Y$ where $Y$ is random variable distributed according a Normal Laplace distribution, that is, $Y = Z + W$, where $Z\sim N\left(\mu,\sigma^2\right)$,
 and $W$ is distributed according a skewed Laplace model, that is, 
\[
 f_{W}(w|\alpha,\beta)=\left\{
\begin{array}{cl}
\frac{\alpha \beta}{\alpha+\beta}e^{\beta w} & \mbox{if $w\leq 0$,} \\
\frac{\alpha \beta}{\alpha+\beta}e^{-\alpha w} & \mbox{if $w>0$,}
\end{array}
\right.
\]
\noindent independent of $Z$, for $\alpha,\beta >0$. 


\cite{Reed} illustrate the form of the \textit{dPlN} density function for various different groups of parameter values. In particular, they show that it exhibits
heavy-tail behavior, since $f_X(x) \rightarrow k x^{-\alpha-1}$
as $x \rightarrow \infty$. The \textit{dPlN} distribution lacks a closed form expression for its moment generating function. Nevertheless, if $r<\alpha$, the moment of order $r$ is given by
\begin{equation*}
E\left(X^r\mid \alpha, \beta, \mu, \sigma^2\right)=\frac{\alpha \beta}{(\alpha-r)(\beta+r)}e^{r\mu + r^2 \sigma^2/2}.
\end{equation*}

\begin{figure}[htb]
\centerline{\includegraphics[height=3in]{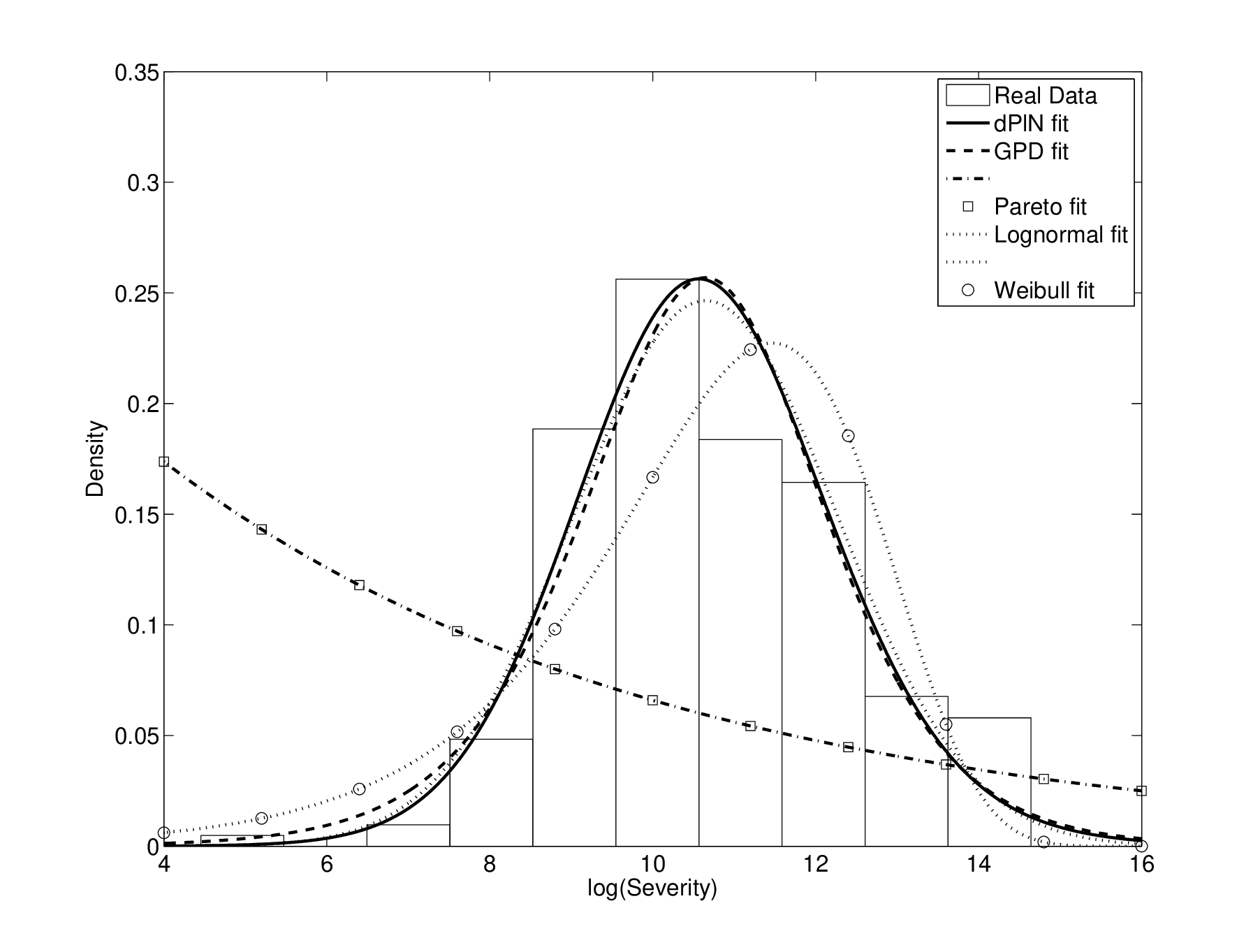}}
\caption{\label{Fig5}{Histogram and fitted pdf for the real severities in log-scale under the \emph{dPlN} model (solid line) and benchmark approaches.}}
\end{figure}

The approach for Bayesian estimation of the \emph{dPlN} distribution in \cite{PepadPlN} was applied to fit the severity data sample. The (posterior) expected values for the parameters of the model were \mbox{$\left(\alpha,\beta,\mu,\sigma^2\right)= (1.24,\ 1.8,\ 10.4,\ 1.29^2)$}. Note that, since $\alpha <2$ then,  the estimated distribution has a finite mean but an infinite variance. {The fit to the histogram of the severities sample (in log-scale) by the \emph{dPlN} distribution is shown by Figure \ref{Fig5}, as well as the fits by other heavy-tailed models typically used in the OpRisk contexts, namely, the Lognormal, Weibull, Pareto and Generalized Pareto. It can be seen how the first three have a poorer performance that the \emph{dPlN}, while the Generalized Pareto was comparable.}

\subsection{The estimated loss aggregate distribution and risk measures}\label{est_compound}
From the estimates obtained in Sections \ref{est_freq} and \ref{est_sev} we make inference here for the loss aggregate model (\ref{risk_model}). As commented in Section \ref{review_loss}, a number of numerical strategies do exist in order to estimate (\ref{risk_model}). Comparing the different approaches is out of the scope in this paper, and here for simplicity, we shall approximate the loss aggregate distribution by the classic, Monte Carlo (MC) method. The steps to implement the MC algorithm are shown in Table \ref{tabMC}.
\begin{table}[h!]
\begin{center}
\fbox{\parbox{\textwidth}{\begin{enumerate}
\item For $k = 1,\ldots,K$,
\begin{enumerate}
\item Simulate the number of annual losses, $N$, from the estimated $\map_2$ given by (\ref{estimated_map}). The estimated mass function of $N$ to be used is depicted by Figure \ref{Fig:mass}.
\item Given that $N = n$, simulate independent severities $X_1,\ldots,X_n$ from the \emph{dPlN} distribution estimated in Section  \ref{est_sev}, with parameters $\left(\alpha,\beta,\mu,\sigma^2\right)= (1.24,\ 1.8,\ 10.4,\ 1.29^2)$.
\item Compute $Z_k = \sum_{i=1}^n X_i$.
\end{enumerate}
\item Set $k = k+1$ and return to 1.
\end{enumerate}
}}
\caption{Monte Carlo algorithm to sample from the loss aggregate distribution. \label{tabMC}}
\end{center}
\end{table}

Once the samples $(Z_1,\ldots,Z_k)$ from the loss aggregate distribution are obtained, the risk measures (\ref{var}) and (\ref{es}) can be easily estimated by their empirical counterparts.

\begin{figure}[htb]
\begin{center}\begin{tabular}{c}
\includegraphics[height=2.6in]{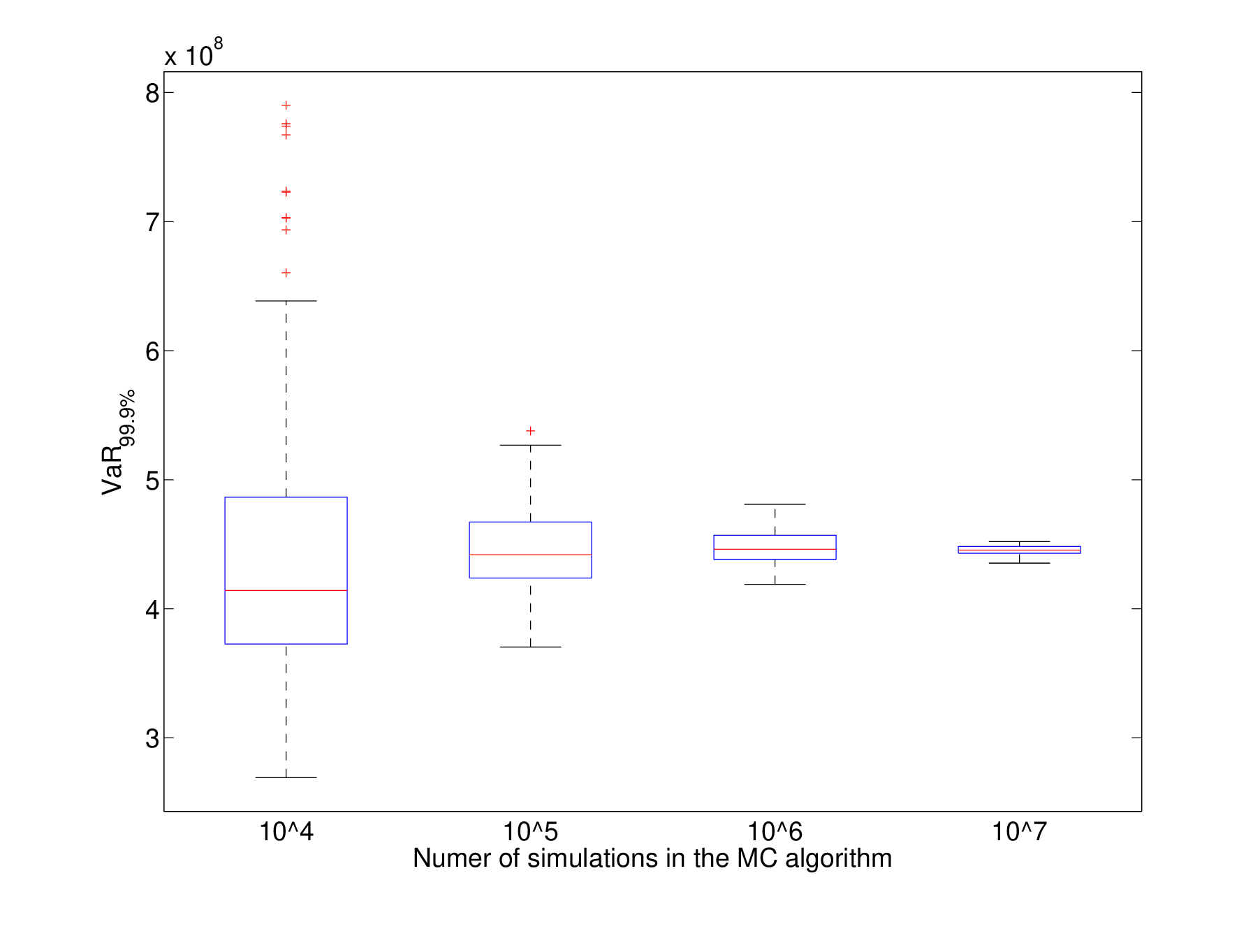}
\end{tabular}
\end{center}
\caption{\label{Fig:box}Boxplot of the samples of $VaR_{99.9\%}$ ($500$ values each), for different values of $K$.}
\end{figure}

One of the main drawbacks of the MC algorithm is its typical slow convergence, specially in presence of heavy tails, see for example \cite{brunner2009fat,Shevchenko2}. In order to choose an appropriate number of simulations $K$, the following numerical exercise was considered. A total of $K$ samples of $Z$, for $K \in \{10^4, 10^5, 10^6,10^7\}$, were obtained $500$ times, from which the corresponding estimated $VaR_{99.9\%}$s were recorded. The $500$ samples of the $VaR_{99.9\%}$ under the four considered values of $K$ are depicted in Figure \ref{Fig:box}. It can be noted that under $K=10^7$, the convergence is achieved.

Section \ref{est_freq} showed the outperformance of the $\map_2$ with respect to the Poisson process when fitting the loss occurrence process. {It is natural to wonder if such encountered differences affect the estimated compound models and related risk measures. To address this issue, the algorithm in Table \ref{tabMC} was implemented, also for $K=10^7$, but here the distribution of the annual losses $N$ in point 1.(a) was set to a Poisson distribution of rate equal to the average number of annual losses ($16.6154$). The obtained results evidence the sensitivity of the aggregate loss distribution to the choice of the occurrence process.} Table \ref{statistics} summarizes both aggregate distributions by means of an assortment of descriptive statistics (minimum and maximum values, mean, standard deviation, skewness and different quantiles). The values have been divided by $10^7$ for abbreviation reasons. A first remark to be made is that the minimum value and the 0.025 quantile of $Z$ under the $\map_2$ are equal to zero. Indeed, there exists a proportion (around $6\%$) of null samples of $Z$. This is a consequence of the non-negligible probability that $N=0$, under the $\map_2$ (see Figure \ref{Fig:mass}). This phenomenon is rarely observed under the considered Poisson process for which $P(N=0)\approx 6\times 10^{-8}$ is obtained. Second, as expected (see \cite{embrechts1997modelling}), both distributions are positive skewed with a long right tail as a consequence of the heavy-tailed severities. In all cases (with the exception of the 0.025 quantile), the quantiles and mean of $Z$ are larger under the $\map_2$ than under the Poisson process. This can be explained again by the mass function of the annual occurrences under the two considered point processes: while for the Poisson process $P(N\geq 30) = 0.002$, this probability changes to $0.2836$, under the $\map_2$. This implies that for some years, a very high number of losses may occur under the $\map_2$, and this makes the loss aggregate distribution take more extreme values than under the Poisson process. 
\renewcommand{\tabcolsep}{0.2cm}
\renewcommand{\arraystretch}{1.3}
\begin{table}[htb]
\begin{center}
\begin{tabular}{|c|r|r|r|r|r|r|r|r|r|S[table-format=3.2,table-number-alignment=right] }
\cline{2-10}
\multicolumn{1}{c|}{}& $\min$ & $\max$ & mean & SD & $\gamma$&$Q_{.025}$ & $Q_{.25}$ &$Q_{.50}$ &  $Q_{.975}$\\
\hline			
$\map_2$		&	0&	$8420$	&	$1.4$	&$16$ & 321.9	&	0	&	0.72 &0.99 &4.47\\
\emph{Poisson}	 &$0.00045$&	$5000$	&$0.41$		&	$8.6$	& 431.6&	0.0525	&	 0.1303 & 0.2073& 1.44 	 	\\ \hline																
\end{tabular}
\end{center}
\caption{Values (divided by $10^7$) for the descriptive statistics of the aggregate loss distribution $Z$ under the $\map_2$ and Poisson process.\label{statistics}}
\end{table}

Finally, Table \ref{risk_measures} shows the risk measures of interest for the financial entities: the $VaR_p$ and $ES_p$ (see Eq. \ref{var}-\ref{es}), for $p\in \{0.99, 0.999\}$ and under the two point processes. The shown values are divided by $10^8$ and $10^9$, respectively. Here again, the results suggest that under the $\map_2$ process, higher potential losses than under the Poisson process may occur, and in consequence, the capital charge increases in significant way. 

\renewcommand{\tabcolsep}{0.2cm}
\renewcommand{\arraystretch}{1.3}
\begin{table}[htb]
\begin{center}
\begin{tabular}{|c|r|r|r|r|S[table-format=3.2,table-number-alignment=right] }
\cline{2-5}
	\multicolumn{1}{l|}{}	& \multicolumn{2}{c|}{$VaR_p$}		&	\multicolumn{2}{c|}{$ES_p$}\\
\cline{2-5}
\multicolumn{1}{c|}{} & $p=0.99$ & $p=0.999$  & $p=0.99$ & $p=0.999$\\
\hline			
$\map_2$		&		0.8068	&	4.405		&	0.3181	&	1.6887 	\\
\emph{Poisson}	 &	0.27	&1.575			&	0.1209	&	 0.613 	 	\\ \hline																
\end{tabular}
\end{center}
\caption{$VaR_p$ and $ES_p$ values (divided by $10^8$ and $10^9$, respectively) under the $\map_2$ and Poisson process, for $p\in \{0.95, 0.99, 0.999\}$\label{risk_measures}.}
\end{table}

%
%
%


\section{Conclusions}\label{sec: discussion}
In this paper, an aggregate loss model with occurrence process governed by a type of Markov renewal process has been considered. The selected stochastic process is the $\map_2$ which allows for correlated and non-exponential inter-loss times and also for overdisperse loss counts. We have shown in Section 3 how the $\map_2$ mostly produces loss counts presenting a variance significantly higher than the mean, a feature that highlights its applicability in financial contexts. One of the novelties of the paper is the derivation of some persistence measures that may help the financial decision makers when evaluating the historical trace of loss times. The aggregate loss model has been estimated by modeling in separate way the frequency from the severity distributions. For the first, an approach based on the direct maximization of the likelihood function of the inter-loss times distributions, has been explored.  The application of the methodology to the OpRisk modeling constitutes a significant contribution of the work. A real OpRisk data set representing inter-loss times and associated severities is fitted: for comparison reasons, the times are modeled by both the $\map_2$ and the Poisson process, and for the severities the same heavy-tailed distribution is considered. The outperformance of the $\map_2$ in fitting the occurrence process leads to important differences in the estimated loss aggregate distribution and related risk measures, which points out the sensitivity of the aggregate loss distribution to the choice of the frequency distribution in this case.


The authors aim to test the potential for modeling aggregate loss distributions of the Batch Markovian arrival process ($\bmap$), an extension of the $\map$ which allows for simultaneous (and correlated) losses, see \cite{Lucantoni.new.results,Lucantoni}. Several perspectives may be considered in this respect. First, and similarly as in the present paper, the $\bmap$ can be thought as a model for the frequency while any type of heavy-tailed distribution is chosen for the severity.  In principle, this approach would show more versatility and applicability than the $\map$ because in real risk scenarios different losses may occur at the same time.  Second, the previous approach could be extended to model aggregate loss distributions more than one cell via the superposition of differently parametrized $\bmap$s. From the statistical inference viewpoint the $\bmap$ represents a partially solved problem, mainly because of its non-identifiability, lack of a known canonical representation and consequent numerical troubles. Finally, as already commented throughout the manuscript, a more detailed inspection of the formulae for the persistence measures should be undertaken. In particular, given the aspect of the mass probability functions for the sequence of spells, it is of interest to study the possible connection with zero-inflated models.  Work on these issues is underway.

%



\section*{Acknowledgements}
This research has been financed in part by research projects FQM-329 and P18-FR-2369 (Junta de Andaluc\'{\i}a, Spain); PID2019-110886RB-I00,  PID2019-104901RB-I00(Ministerio de Econom\'{\i}a, Industria y Competitividad, Spain); PR2019-029 (Universidad de C\'adiz, Spain); This support is gratefully acknowledged.
\section*{Appendix A: Derivation of the expressions for $p_{01}(s)$ and $p_{11}(s)$}
The proof is a consequence of the theory of Markov renewal processes. All results used in this Appendix can be found in Ch. 10 in \cite{Cinlar.stochastic}. 

First, $p_{01}(s)$ is written as
\begin{equation}\label{p01_2}
p_{01}(s) = \frac{P(T_{n+1}>s, T_{n}<s)}{P(T_{n}<s)},
\end{equation}
where according to (\ref{cdf_ph}), $P(T_{n}<s)=F_T(s)=1-\phinegrita e^{D_0s}\enegrita$. Recall that since the $\map_2$ is a Markov renewal process, then the sequences of states $\{Y_{n}\}_{n=1}^\infty$ at the times of the losses
is a Markov chain with transition matrix $P^\star$. Consider the semi-Markov kernel $Q(\cdot)$ as in (\ref{semiMk}) over the state space $\mathcal{S}=\{1,2\}$. Then, by definition of semi-Markov kernel, for $i,j\in \mathcal{S}$, $t>0$,
$$
Q_{i,j}(t)=P(Y_{n+1}=j, T_{n+1}\leq t \mid Y_n = i),
$$
which represents the $(i,j)$-th element of the semi-Markov kernel $Q(t)$. Also, define 
$$
G(i,j,t)=\frac{Q_{i,j}(t)}{P^\star(i,j)},\ i,j\in \mathcal{S}, t>0,
$$
where $G(i,j,t)=1$ in case $P^\star(i,j)=Q_{i,j}(t)=0$. Then, $G(i,j,\cdot)$ is a cumulative distribution function
$$
G(i,j,t)=P(T_{n+1}\leq t \mid Y_n=i,Y_{n+1}=j).
$$
Finally, the following paramount property is key for our proof. This states that given the Markov chain, then the inter-loss times are conditionally independent, or in other words,
$$
P(T_1 \leq u_1,\ldots, T_n \leq u_n \mid Y_0,\ldots,Y_n)=G(Y_0,Y_1,u_1)G(Y_1,Y_2,u_2)\ldots G(Y_{n-1},Y_n,u_n).
$$
Consider the numerator in (\ref{p01_2}). This can be rewritten as

\begin{eqnarray}\label{joint}
&&P(T_n<s, T_{n+1}>s)=\\\nonumber
&&\sum_{i,j,k \in \{1,2\}} \bigl\{P(T_n<s, T_{n+1}>s\mid Y_{n-1}=i,Y_{n}=j,Y_{n+1}=k)\times\bigr.\\\nonumber
&&\hspace{1.7cm}\bigl.\times P(Y_{n-1}=i,Y_{n}=j,Y_{n+1}=k)\bigr\}=\\\nonumber
&& \sum_{i,j,k \in \{1,2\}} \bigl\{P(T_n<s, T_{n+1}>s\mid Y_{n-1}=i,Y_{n}=j,Y_{n+1}=k)\times\bigr.\\\nonumber
&&\hspace{1.7cm}\bigl.\times P(Y_{n+1}=k\mid Y_n=j,Y_{n-1}=i)P(Y_n=j,Y_{n-1}=i)\bigr\}=\\\nonumber
&& \sum_{i,j,k \in \{1,2\}} \bigl\{P(T_n<s, T_{n+1}>s\mid Y_{n-1}=i,Y_{n}=j,Y_{n+1}=k)\times\bigr.\\\nonumber
&&\hspace{1.7cm}\bigl.\times P(Y_{n+1}=k\mid Y_n=j)P(Y_n=j\mid Y_{n-1}=i)P(Y_{n-1}=i)\bigr\}=\\\nonumber
&& \sum_{i,j,k \in \{1,2\}} \bigl\{P(T_n<s, T_{n+1}>s\mid Y_{n-1}=i,Y_{n}=j,Y_{n+1}=k)\times\bigr.\\\nonumber
&&\hspace{1.7cm}\bigl.\times P^\star(i,j)\ P^\star(j,k)\ \phi_i\bigr\}=\\\nonumber
&& \sum_{i,j,k \in \{1,2\}}G\left(i,j,s\right) \left(1-G(j,k,s) \right) P^\star(i,j)\ P^\star(j,k)\ \phi_i=\\\nonumber
&&\sum_{i,j,k \in \{1,2\}}\frac{Q_{i,j}(s)}{P^\star(i,j)} \left(1-\frac{Q_{j,k}(s)}{P^\star(j,k)} \right)P^\star(i,j)\ P^\star(j,k)\ \phi_i=\\\nonumber
&& \sum_{i,j,k \in \{1,2\}}\frac{Q_{i,j}(s)}{P^\star(i,j)} \left(\frac{P^\star(j,k)-Q_{j,k}(s)}{P^\star(j,k)} \right)P^\star(i,j)\ P^\star(j,k)\ \phi_i=\\ 
&& \sum_{i,j,k \in \{1,2\}}Q_{i,j}(s) \bigl(P^\star(j,k)-Q_{j,k}(s) \bigr) \phi_i.
\end{eqnarray}\label{sinmatriz}
It is a straightforward computation to check that (\ref{sinmatriz}) can be expressed in matrix form as
\begin{equation}
\sum_{i,j,k \in \{1,2\}}Q\left(i,j,s\right) \bigl(P^\star(j,k)-Q(j,k,s) \bigr) \phi_i= \phinegrita Q(s)\bigl(P^\star-Q(s)\bigr)\enegrita.
\end{equation}
From (\ref{semiMk}) and (\ref{Pstar}) one obtains
$$
\phinegrita Q(s)\bigl(P^\star-Q(s)\bigr)\enegrita = \phinegrita\left(I-e^{D_0 s}\right) P^\star e^{D_0 s}P^\star \enegrita,
$$
which concludes the proof for the expression of $p_{01}$. Finally, the proof for the matrix formulation of $p_{11}$ is obtained in analogous way.

\section*{Appendix B: Derivation of the expressions for $P(S=n)$ and $P(L=n)$}
Consider first the case $n=0$. Then, from (\ref{cdf_ph})
$$
P(S=0)= P(T_{n+1}>s) = 1-F_T(s)=\phinegrita e^{D_0s}\enegrita,
$$
and similarly for the variable $L$. If instead $n\geq1$, the proof follows closely the derivation of (\ref{joint}). Indeed, by a recursive argument,\\

\begin{eqnarray*}\label{proofShortspell}
\hspace{-1cm}P(S = n) &=& P(T_1<s, T_2<s,\ldots,T_n<s, T_{n+1}>s)=\\
&&\hspace{-3.5cm}\sum_{i_1,i_2,\ldots,i_{n+2} \in \{1,2\}} \bigl\{P\left(T_1<s, T_2<s,\ldots,T_n<s, T_{n+1}>s\mid Y_{0}=i_1,Y_{1}=i_2,\ldots,Y_n=i_{n+1},Y_{n+1}=i_{n+2}\right)\times\bigr.\\\nonumber
&&\hspace{-1cm}\bigl.\times P(Y_{0}=i_1,Y_{1}=i_2,\ldots,Y_n=i_{n+1},Y_{n+1}=i_{n+2})\bigr\}=\\\nonumber
\hspace{-2cm}&&\ldots=\\ 
\hspace{-1cm}&&\hspace{-3.5cm}\sum_{i_1,i_2,\ldots,i_{n+2} \in \{1,2\}}G\left(i_1,i_2,s\right)G\left(i_2,i_3,s\right)\cdots
G\left(i_n,i_{n+1},s\right)\bigl(1-G(i_{n+1},i_{n+2},s) \bigr)P^\star(i_{1},i_{2})\cdots P^\star(i_{n+1},i_{n+2})\phi_{i_1}=\\
\hspace{-1cm}&&\hspace{-3.5cm}\sum_{i_1,i_2,\ldots,i_{n+2} \in \{1,2\}}\bigl(\prod_{l=1}^{n} G\left(i_l,i_{l+1},s\right)\bigr) \bigl(1-G(i_{n+1},i_{n+2},s) \bigr) \bigl(\prod_{l=1}^{n} P^\star\left(i_l,i_{l+1}\right)\bigr)\ P^\star(i_{n+1},i_{n+2})\ \phi_{i_1}=\\
\hspace{-1cm}&&\hspace{-3.5cm}\sum_{i_1,i_2,\ldots,i_{n+2} \in \{1,2\}}\bigl(\prod_{l=1}^{n} Q_{i_l,i_{l+1}}(s)\bigr) \ \bigl(P^\star(i_{n+1},i_{n+2})-Q_{i_{n+1},i_{n+2}}(s) \bigr) \phi_{i_1}.
\end{eqnarray*}
Then,
$$
\sum_{i_1,i_2,\ldots,i_{n+2} \in \{1,2\}}\bigl(\prod_{l=1}^{n} Q_{i_l,i_{l+1}}(s)\bigr) \ \bigl(P^\star(i_{n+1},i_{n+2})-Q_{i_{n+1},i_{n+2}}(s) \bigr) \phi_{i_1}= \phinegrita Q^n(s)\bigl(P^\star-Q(s)\bigr)\enegrita,
$$
and from  (\ref{semiMk}) and (\ref{Pstar}),
$$
\phinegrita Q^n(s)\bigl(P^\star-Q(s)\bigr)\enegrita=\phinegrita \left[\left( I-e^{D_0s}    \right)P^\star  \right]^n e^{D_0 s} P^\star \enegrita.
$$
To see that (\ref{Ps}) is indeed a mass probability function, we next show that 
$\sum_{n=0}^{\infty}P(S=n)=1$. First, from Ch. 4 in \cite{Janssen}, one has
$$
\lim_{n\rightarrow \infty}Q^n(s)=\mathbf{0},
$$
which implies that $\sum_{n=1}^\infty Q^n(s)=Q(s)(I-Q(s))^{-1}$. Then, 
\begin{eqnarray*}
\sum_{n=0}^{\infty}P(S=n) &=& \phinegrita e^{D_0s}\enegrita+\sum_{n=1}^{\infty}\phinegrita Q^n(s)e^{D_0s}P^\star\enegrita\\
&=& \phinegrita e^{D_0s}\enegrita+\phinegrita Q(s)(I-Q(s))^{-1}e^{D_0s}P^\star\enegrita\\
&=& \phinegrita e^{D_0s}\enegrita+\phinegrita Q(s)(I-Q(s))^{-1}(P^\star-Q(s))\enegrita\\
&=& \phinegrita e^{D_0s}\enegrita+\phinegrita Q(s)(I-Q(s))^{-1}(P^\star \enegrita-Q(s)\enegrita)\\
&=& \phinegrita e^{D_0s}\enegrita+\phinegrita Q(s)(I-Q(s))^{-1}(\enegrita-Q(s)\enegrita)\\
&=& \phinegrita e^{D_0s}\enegrita+\phinegrita Q(s)(I-Q(s))^{-1}(I-Q(s))\enegrita\\
&=& \phinegrita e^{D_0s}\enegrita+\phinegrita Q(s)\enegrita\\
&=& \phinegrita \left(e^{D_0s}+P^\star-  e^{D_0s}  P^\star\right)\enegrita\\
&=& \phinegrita e^{D_0s}\enegrita+\phinegrita P^\star\enegrita-\phinegrita e^{D_0s}P^\star\enegrita\\
&=& \phinegrita e^{D_0s}\enegrita+1-\phinegrita e^{D_0s}\enegrita\\
&=&1
\end{eqnarray*}
In the previous set of formulae we applied that since $P^\star$ is a stochastic matrix, then $P^\star \enegrita=\enegrita$. Also, since $\phinegrita$ represents its stationary probability vector, then 
$\phinegrita P^\star = \phinegrita$.

Finally, the proof for $P(L=n)$ is immediate following a similar reasoning.

\bibliographystyle{myapalike.bst}
\bibliography{ref_paper}

\begin{thebibliography}{}

\bibitem[Ahn and Badescu, 2007]{ahn2007analysis}
Ahn, S. and Badescu, A.~L. (2007).
\newblock On the analysis of the {G}erber--{S}hiu discounted penalty function
  for risk processes with {M}arkovian arrivals.
\newblock {\em Insurance: Mathematics and Economics}, 41(2):234--249.

\bibitem[Aus{\'\i}n et~al., 2011]{ausin2011bayesian}
Aus{\'\i}n, M., Vilar, J., Cao, R., and Gonz{\'a}lez-Fragueiro, C. (2011).
\newblock Bayesian analysis of aggregate loss models.
\newblock {\em Mathematical Finance}, 21(2):257--279.

\bibitem[Avanzi et~al., 2016]{overdispersion2}
Avanzi, B., Wong, B., and Yang, X. (2016).
\newblock A micro-level claim count model with overdispersion and reporting
  delays.
\newblock {\em Insurance: Mathematics and Economics}, 71:1 -- 14.

\bibitem[Badescu et~al., 2007]{badescu2007analysis}
Badescu, A., Drekic, S., and Landriault, D. (2007).
\newblock On the analysis of a multi-threshold {M}arkovian risk model.
\newblock {\em Scandinavian Actuarial Journal}, 2007(4):248--260.

\bibitem[Bodrog et~al., 2008]{Bodrog}
Bodrog, L., Heindlb, A., Horv\'ath, G., and Telek, M. (2008).
\newblock A {M}arkovian canonical form of second-order matrix-exponential
  processes.
\newblock {\em European Journal of Operational Research}, 190:459--477.

\bibitem[Bolanc{\'e} et~al., 2012]{bolance2012quantitative}
Bolanc{\'e}, C., Guill{\'e}n, M., Gustafsson, J., and Nielsen, J.~P. (2012).
\newblock {\em Quantitative Operational Risk Models}.
\newblock Chapmal \& Hall, CRC Press.

\bibitem[Brechmann et~al., 2014]{Brechmann}
Brechmann, E., Czado, C., and Paterlini, S. (2014).
\newblock Flexible dependence modeling of operational risk losses and its
  impact on total capital requirements.
\newblock {\em Journal of Banking \& Finance}, 40:271--285.

\bibitem[Breuer, 2002]{Breuer}
Breuer, L. (2002).
\newblock An {EM} algorithm for batch {M}arkovian arrival processes and its
  comparison to a simpler estimation procedure.
\newblock {\em Annals of Operations Research}, 112:123--138.

\bibitem[Brunner et~al., 2009]{brunner2009fat}
Brunner, M., Piacenza, F., Monti, F., and Bazzarello, D. (2009).
\newblock Fat tails, expected shortfall and the {M}onte {C}arlo method: a note.
\newblock {\em The Journal of Operational Risk}, 4(1):81--88.

\bibitem[Carrizosa et~al., 2014]{Carrizosa1}
Carrizosa, E., Jockovi\'c, J., and Ram\'irez-Cobo, P. (2014).
\newblock A global optimization approach for parameter estimation of a mixture
  of double pareto lognormal and lognormal distributions.
\newblock {\em Computers \& Operations Research}, 52:231--240.

\bibitem[Carrizosa and Ram\'irez-Cobo, 2014]{carrizosa2014maximum}
Carrizosa, E. and Ram\'irez-Cobo, P. (2014).
\newblock Maximum likelihood estimation in the two-state {Markovian} arrival
  process.

\bibitem[Casale et~al., 2010]{Casale}
Casale, G., Z.~Zhang, E., and Simirni, E. (2010).
\newblock Trace data characterization and fitting for {M}arkov modeling.
\newblock {\em Performance Evaluation}, 67:61--79.

\bibitem[\c{C}inlar, 1975]{Cinlar.stochastic}
\c{C}inlar, E. (1975).
\newblock {\em Introduction to stochastic processes}.
\newblock Prentice-Hall, Usa.

\bibitem[Chakravarthy, 2001]{Chakravarthy}
Chakravarthy, S. (2001).
\newblock The {Batch Markovian} arrival process: a review and future work.
\newblock In et~al., A.~K., editor, {\em Advances in probability and stochastic
  processes}, pages 21--49.

\bibitem[Chakravarthy, 2009]{AMC1}
Chakravarthy, S. (2009).
\newblock A disaster queue with {Markovian} arrivals and impatient customers.
\newblock {\em Applied Mathematics and Computation}, 214:48--59.

\bibitem[Chapelle et~al., 2004]{Chapelle}
Chapelle, A., Crama, Y., Hubner, G., and Peters, J. (2004).
\newblock Basel {II} and operational risk: Implications for risk measurement
  and management in the financial sector.
\newblock National Bank of Belgium Working Paper, No. 51.

\bibitem[Chavez-Demoulin et~al., 2015]{chavez2015extreme}
Chavez-Demoulin, V., Embrechts, P., and Hofert, M. (2015).
\newblock An extreme value approach for modeling operational risk losses
  depending on covariates.
\newblock {\em Journal of Risk and Insurance}.

\bibitem[Chavez-Demoulin et~al., 2006]{chavez2006quantitative}
Chavez-Demoulin, V., Embrechts, P., and Ne{\v{s}}lehov{\'a}, J. (2006).
\newblock Quantitative models for operational risk: extremes, dependence and
  aggregation.
\newblock {\em Journal of Banking \& Finance}, 30(10):2635--2658.

\bibitem[Chernobai et~al., 2008]{Chernobai1}
Chernobai, A.~S., Rachev, S.~T., and Fabozzi, F.~J. (2008).
\newblock {\em Operational risk: a guide to Basel II capital requirements,
  models, and analysis}, volume 180.
\newblock John Wiley \& Sons.

\bibitem[Cheung and Landriault, 2010]{Cheung}
Cheung, E. and Landriault, D. (2010).
\newblock A generalized penalty function with the maximum surplus prior to ruin
  in a {MAP} risk model.
\newblock {\em Insurance: {M}athematics and {E}conomics}, 46:127--134.

\bibitem[Cheung et~al., 2018]{AMC3}
Cheung, E., Liu, H., and Willmot, G. (2018).
\newblock Joint moments of the total discounted gains and losses in the renewal
  risk model with two-sided jumps.
\newblock {\em Applied Mathematics and Computation}, 331:358--377.

\bibitem[Cheung et~al., 2009]{cheung2009perturbed}
Cheung, E.~C., Landriault, D., et~al. (2009).
\newblock Perturbed {MAP} risk models with dividend barrier strategies.
\newblock {\em Journal of Applied Probability}, 46(2):521--541.

\bibitem[Cope and Antonini, 2008]{cope2008observed}
Cope, E. and Antonini, G. (2008).
\newblock Observed correlations and dependencies among operational losses in
  the {ORX} consortium database.
\newblock {\em Journal of Operational Risk}, 3(4):47--74.

\bibitem[Cruz, 2002]{Cruz}
Cruz, M. (2002).
\newblock {\em Modeling, Measuring and Hedging Operational Risk}.
\newblock John Wiley \& Sons.

\bibitem[Dimitrova et~al., 2016]{AMC6}
Dimitrova, D., Kaishev, V., and Zhao, S. (2016).
\newblock On the evaluation of finite-time ruin probabilities in a dependent
  risk model.
\newblock {\em Applied Mathematics and Computation}, 275:268--286.

\bibitem[Dutta and Perry, 2007]{dutta2006tale}
Dutta, K. and Perry, J. (2007).
\newblock A tale of tails: an empirical analysis of loss distribution models
  for estimating operational risk capital.
\newblock Working paper 06-13, Federal Reserve Bank of Boston.

\bibitem[Embrechts et~al., 2003]{Embrechts2}
Embrechts, P., H\"{o}ing, A., and Juri, A. (2003).
\newblock Using copulae to bound the {Value-at-Risk} for functions of dependent
  risks.
\newblock {\em Finance and Stochastics}, 7:145--167.

\bibitem[Embrechts et~al., 1997]{embrechts1997modelling}
Embrechts, P., Kl{\"u}ppelberg, C., and Mikosch, T. (1997).
\newblock {\em Modelling extremal events: for insurance and finance},
  volume~33.
\newblock Springer Science \& Business Media.

\bibitem[Eum et~al., 2007]{Eum}
Eum, S., Harris, R., and Atov, I. (2007).
\newblock A matching model for {MAP}-2 using moments of the counting process.
\newblock In {\em {Proceedings of the International Network Optimization
  Conference, INOC 2007}}, Spa, Belgium.

\bibitem[Fearnhead and Sherlock, 2006]{fearnhead2006exact}
Fearnhead, P. and Sherlock, C. (2006).
\newblock An exact {Gibbs sampler for the Markov-modulated Poisson} process.
\newblock {\em Journal of the Royal Statistical Society: Series B (Statistical
  Methodology)}, 68(5):767--784.

\bibitem[Feria-Dom{\'\i}nguez et~al., 2015]{overdispersion}
Feria-Dom{\'\i}nguez, J., Jim\'enez-Rodr\'iguez, E., and Sholarin, O. (2015).
\newblock Tackling the over-dispersion of operational risk: implications on
  capital adequacy requirements.
\newblock {\em The North American Journal of Economics and Finance},
  31:206--221.

\bibitem[Fodra and Pham, 2015]{fodra2015high}
Fodra, P. and Pham, H. (2015).
\newblock High frequency trading and asymptotics for small risk aversion in a
  markov renewal model.
\newblock {\em SIAM Journal on Financial Mathematics}, 6(1):656--684.

\bibitem[Frachot et~al., 2001]{frachot2001loss}
Frachot, A., Georges, P., and Roncalli, T. (2001).
\newblock Loss distribution approach for operational risk.
\newblock Available at http://dx.doi.org/10.2139/ssrn.1032523.

\bibitem[Frostig, 2008]{frostig2008ruin}
Frostig, E. (2008).
\newblock On ruin probability for a risk process perturbed by a {L\'e}vy
  process with no negative jumps.
\newblock {\em Stochastic Models}, 24(2):288--313.

\bibitem[Fung et~al., 2019]{fung2019multivariate}
Fung, T., Badescu, A., and Lin, X. (2019).
\newblock {Multivariate Cox hidden Markov} models with an application to
  operational risk.
\newblock {\em Scandinavian Actuarial Journal}, 2019(8):686--710.

\bibitem[Gerhold et~al., 2010]{Gerhold}
Gerhold, S., Schmock, U., and Warnung, R. (2010).
\newblock A generalization of {P}anger's recursion and numerically stable risk
  aggregation.
\newblock {\em Finance and Stochastics}, 14:81--128.

\bibitem[{Groupe Consultatif Actuariel Europ\'een}, 2007]{anexo_european}
{Groupe Consultatif Actuariel Europ\'een} (2007).
\newblock {Solvency II Glossary: European Commission}.
\newblock Technical report.

\bibitem[Heffes and Lucantoni, 1986]{Heffes.packetized}
Heffes, H. and Lucantoni, D. (1986).
\newblock A {Markov} modulated characterization of packetized voice and data
  traffic and related statistical multiplexer performance.
\newblock {\em IEEE Journal on Selected Areas in Communications}, 4:856--868.

\bibitem[Janssen and Manca, 2006]{Janssen}
Janssen, J. and Manca, R. (2006).
\newblock {\em {Applied Semi-Markov Processes}}.
\newblock Springer.

\bibitem[Kim et~al., 2017]{AMC4}
Kim, C., Klimenok, V., and Dudin, A. (2017).
\newblock Analysis of unreliable {BMAP/PH/N} type queue with {Markovian} flow
  of breakdowns.
\newblock {\em Applied Mathematics and Computation}, 314:154--172.

\bibitem[Klemm et~al., 2003]{Klemm}
Klemm, A., Lindemann, C., and Lohmann, M. (2003).
\newblock Modeling {IP} traffic using {Batch Markovian Arrival Process}.
\newblock {\em Performance Evaluation}, 54(2):149--173.

\bibitem[Klugman et~al., 2012]{klugman2012loss}
Klugman, S.~A., Panjer, H.~H., and Willmot, G.~E. (2012).
\newblock {\em Loss models: from data to decisions}.
\newblock John Wiley \& Sons.

\bibitem[Lindsey, 1995]{lindsey}
Lindsey, J. (1995).
\newblock {\em Modelling frequency and count data}, volume~15.
\newblock Oxford University Press.

\bibitem[Lucantoni, 1991]{Lucantoni.new.results}
Lucantoni, D. (1991).
\newblock New results for the single server queue with a {Batch Markovian
  Arrival Process}.
\newblock {\em Stochastic Models}, 7:1--46.

\bibitem[Lucantoni, 1993]{Lucantoni}
Lucantoni, D. (1993).
\newblock The {$BMAP/G/1$} queue: A tutorial.
\newblock In Donatiello, L. and Nelson, R., editors, {\em Models and Techniques
  for Performance Evaluation of Computer and Communication Systems}, pages
  330--358. Springer, New York.

\bibitem[Lucantoni et~al., 1990]{Lucantoni90}
Lucantoni, D., Meier-Hellstern, K., and Neuts, M. (1990).
\newblock A single-server queue with server vacations and a class of nonrenewal
  arrival processes.
\newblock {\em Advances in Applied Probability}, 22:676--705.

\bibitem[Maegebier, 2013]{maegebier2013valuation}
Maegebier, A. (2013).
\newblock Valuation and risk assessment of disability insurance using a
  discrete time trivariate {M}arkov renewal reward process.
\newblock {\em Insurance: Mathematics and Economics}, 53(3):802--811.

\bibitem[McNeil et~al., 2015]{Mcneil}
McNeil, A.~J., Frey, R., and Embrechts, P. (2015).
\newblock {\em Quantitative Risk Management: Concepts, Techniques and Tools}.
\newblock Princeton university press.

\bibitem[Mittnik and Yener, 2009]{mittnik2009estimating}
Mittnik, S. and Yener, T. (2009).
\newblock Estimating operational risk capital for correlated, rare events.
\newblock {\em Journal of Operational Risk}, 4(4):1--23.

\bibitem[Narayana and Neuts, 1992]{narayana}
Narayana, S. and Neuts, M.~F. (1992).
\newblock The first two moment matrices of the counts for the {M}arkovian
  arrival process.
\newblock {\em Communications in statistics. Stochastic models}, 8(3):459--477.

\bibitem[Nasr et~al., 2018]{Nasr}
Nasr, W., Charanek, A., and Maddah, B. (2018).
\newblock {MAP} fitting by count and inter-arrival moment matching.
\newblock {\em Stochastic Models (in Press)}.

\bibitem[Neuts and Li., 1997]{NeutsLi}
Neuts, M. and Li., J. (1997).
\newblock {\em An algorithm for the {$P(n,t)$} matrices of a continuous
  {BMAP}}, volume 183 of {\em Lectures notes in Pure and Applied Mathematics},
  pages 7--19.
\newblock Srinivas R. Chakravarthy and Attahiru, S. Alfa, editors. NY: Marcel
  Dekker, Inc.

\bibitem[Neuts, 1979]{Neuts79}
Neuts, M.~F. (1979).
\newblock A versatile {M}arkovian point process.
\newblock {\em Journal of Applied Probability}, 16:764--779.

\bibitem[Ng and Yang, 2006]{ng2006joint}
Ng, A.~C. and Yang, H. (2006).
\newblock On the joint distribution of surplus before and after ruin under a
  {M}arkovian regime switching model.
\newblock {\em Stochastic Processes and their Applications}, 116(2):244--266.

\bibitem[Nystrom and Skoglund, 2002]{nystrom2002quantitative}
Nystrom, K. and Skoglund, J. (2002).
\newblock Quantitative operational risk management.
\newblock Working paper, {S}wedbank, {Group Financial Risk Control}.

\bibitem[Okamura and Dohi, 2016]{okamura2016fitting}
Okamura, H. and Dohi, T. (2016).
\newblock Fitting phase-type distributions and {Markovian} arrival processes:
  Algorithms and tools.
\newblock In {\em Principles of performance and reliability modeling and
  evaluation}, pages 49--75.

\bibitem[Panjer, 2006a]{PanjerChapter}
Panjer, H.~H. (2006a).
\newblock {\em Aggregate Loss Modeling}.
\newblock John Wiley \& Sons, Ltd.

\bibitem[Panjer, 2006b]{panjer2006operational}
Panjer, H.~H. (2006b).
\newblock {\em Operational risk: modeling analytics}.
\newblock John Wiley \& Sons.

\bibitem[Peters and Shevchenko, 2015]{peters2015advances}
Peters, G.~W. and Shevchenko, P.~V. (2015).
\newblock {\em Advances in Heavy Tailed Risk Modeling: A Handbook of
  Operational Risk}.
\newblock John Wiley \& Sons.

\bibitem[Peters et~al., 2011]{peters2011analytic}
Peters, G.~W., Shevchenko, P.~V., Young, M., and Yip, W. (2011).
\newblock Analytic loss distributional approach models for operational risk
  from the $\alpha$-stable doubly stochastic compound processes and
  implications for capital allocation.
\newblock {\em Insurance: Mathematics and Economics}, 49(3):565--579.

\bibitem[Pfeifer and Ne{\v{s}}lehov{\'a}, 2004]{pfeifer2004modeling}
Pfeifer, D. and Ne{\v{s}}lehov{\'a}, J. (2004).
\newblock Modeling and generating dependent risk processes for {IRM} and {DFA}.
\newblock {\em Astin Bulletin}, 34(02):333--360.

\bibitem[Ramaswami, 1990]{Ramaswami90}
Ramaswami, V. (1990).
\newblock From the matrix-geometric to the matrix-exponential.
\newblock {\em Queueing Systems}, 6:229--260.

\bibitem[Ram\'irez-Cobo et~al., 2010]{Ramirez}
Ram\'irez-Cobo, P., Lillo, R., and Wiper, M. (2010).
\newblock Nonidentifiability of the two-state {Markovian} arrival process.
\newblock {\em Journal of Applied Probability}, 47(3):630--649.

\bibitem[Ram\'irez-Cobo et~al., 2017]{pepaBayes}
Ram\'irez-Cobo, P., Lillo, R., and Wiper, M. (2017).
\newblock Bayesian analysis of the stationary {MAP}$_2$.
\newblock {\em Bayesian Analysis}, 12(4):1163--1194.

\bibitem[Ram\'{i}rez-Cobo et~al., 2010]{PepadPlN}
Ram\'{i}rez-Cobo, P., Lillo, R.~E., Wilson, S., and Wiper, M.~P. (2010).
\newblock Bayesian inference for {D}ouble {P}areto lognormal queues.
\newblock {\em Annals of Applied Statistics}, 4(3):1533--1557.

\bibitem[Ram\'irez-Cobo et~al., 2014]{hydro}
Ram\'irez-Cobo, P., Marzo, X., Olivares-Nadal, A.~V., Francoso, J.-A.,
  Carrizosa, E., and Pita, M.~F. (2014).
\newblock The {M}arkovian arrival process: A statistical model for daily
  precipitation amounts.
\newblock {\em Journal of Hydrology}, 510(0):459 -- 471.

\bibitem[Reed and Jorgensen, 2004]{Reed}
Reed, W.~J. and Jorgensen, M. (2004).
\newblock The {D}ouble {P}areto-lognormal distribution - a new parametric model
  for size distributions.
\newblock {\em Communications in Statistics, Theory and Methods},
  33(8):1733--1753.

\bibitem[Ren, 2012]{ren2012multivariate}
Ren, J. (2012).
\newblock A multivariate aggregate loss model.
\newblock {\em Insurance: Mathematics and Economics}, 51(2):402--408.

\bibitem[Reshetar, 2008]{Reshetar08}
Reshetar, G. (2008).
\newblock Dependence of operational losses and the capital at risk.
\newblock Available at http://dx.doi.org/10.2139/ssrn.1081256.

\bibitem[Rodr\'iguez et~al., 2015]{joanna2}
Rodr\'iguez, J., Lillo, R., and Ram\'irez-Cobo, P. (2015).
\newblock Failure modeling of an electrical {N}-component framework by the
  non-stationary {M}arkovian arrival process.
\newblock {\em Reliability Engineering and System Safety}, 134:126--133.

\bibitem[Ryd\'en, 1996]{Ryden96b}
Ryd\'en, T. (1996).
\newblock An {EM} algorithm for estimation in {Markov-modulated Poisson}
  processes.
\newblock {\em Computational Statistics and Data Analysis}, 21:431--447.

\bibitem[Scott, 1999]{scott1999bayesian}
Scott, S. (1999).
\newblock Bayesian analysis of a two-state {Markov modulated Poisson} process.
\newblock {\em Journal of Computational and Graphical Statistics},
  8(3):662--670.

\bibitem[Shao et~al., 2017]{AMC5}
Shao, J., Papaioannou, A., and Pantelous, A. (2017).
\newblock Pricing and simulating catastrophe risk bonds in a markov-dependent
  environment.
\newblock {\em Applied Mathematics and Computation}, 309:68--84.

\bibitem[Sharma, 2020]{Bimaquest}
Sharma, S. (2020).
\newblock Operational risk modeling - {A}pproaches and responses.
\newblock {\em Bimaquest}, 20(1):15--31.

\bibitem[Shevchenko, 2010a]{Shevchenko2}
Shevchenko, P.~V. (2010a).
\newblock Calculation of aggregate loss distribution.
\newblock {\em The journal of Operational Risk}, 5(2):3--40.

\bibitem[Shevchenko, 2010b]{Shevchenko1}
Shevchenko, P.~V. (2010b).
\newblock Implementing loss distribution approach for operational risk.
\newblock {\em Applied Stochastic Models in Business and Industry},
  26(3):277--307.

\bibitem[Stutzer, 2020]{stutzer2020persistence}
Stutzer, M. (2020).
\newblock Persistence of averages in financial {Markov Switching models: A}
  large deviations approach.
\newblock {\em Physica A: Statistical Mechanics and its Applications}, page
  124237.

\bibitem[Telek and Horv\'ath, 2007]{Telek}
Telek, M. and Horv\'ath, G. (2007).
\newblock A minimal representation of {M}arkov arrival processes and a moments
  matching method.
\newblock {\em Performance evaluation}, 64:1153--1168.

\bibitem[Valle and Giudici, 2008]{valle}
Valle, L. and Giudici, P. (2008).
\newblock A {B}ayesian approach to estimate the marginal loss distributions in
  operational risk management.
\newblock {\em Computational Statistics $\&$ Data Analysis}, 52:3107--3127.

\bibitem[Vallois and Tapiero, 2009]{vallois2009claims}
Vallois, P. and Tapiero, C. (2009).
\newblock A claims persistence process and insurance.
\newblock {\em Insurance: Mathematics and Economics}, 44(3):367--373.

\bibitem[Xu et~al., 2019]{double}
Xu, C., Zheng, C., Wang, D., Ji, J., and Wang, N. (2019).
\newblock Double correlation model for operational risk: evidence from
  {C}hinese commerical banks.
\newblock {\em Physica A: Statistical mechanics and its applications},
  516:327--339.

\bibitem[Yera et~al., 2019]{yoelEjor}
Yera, Y., Lillo, R., and Ram\'irez-Cobo, P. (2019).
\newblock Fitting procedure for the two-state {Batch Markov modulated Poisson}
  process.
\newblock {\em European Journal of Operational Research}, (279(1)):79--92.

\bibitem[Zhang et~al., 2011]{zhang2011absolute}
Zhang, Z., Yang, H., and Yang, H. (2011).
\newblock On the absolute ruin in a {MAP} risk model with debit interest.
\newblock {\em Advances in Applied Probability}, pages 77--96.

\end{thebibliography}







\end{document}